\documentclass[aps,prr,floatfix,preprintnumbers,altaffilletter,superscriptaddress,reprint]{revtex4-1}
\usepackage{graphicx,url,amssymb,amsmath,rotating,color,units,wasysym,epsfig,multirow,epstopdf,subcaption}
\usepackage[colorlinks,urlcolor=blue,citecolor=blue,linkcolor=blue]{hyperref}
\usepackage{soul,xcolor}

\newcommand{\Beta}{\mathrm{B}}
\begin{document}
\title{Gravitational-wave astronomy with an uncertain noise power spectral density}

\author{Colm Talbot}
\email{ctalbot@caltech.edu}
\affiliation{LIGO Laboratory, California Institute of Technology, Pasadena, CA 91125, USA}
\affiliation{Monash Centre for Astrophysics, School of Physics and Astronomy, Monash University, VIC 3800, Australia}
\affiliation{OzGrav: The ARC Centre of Excellence for Gravitational-Wave Discovery, Clayton, VIC 3800, Australia}

\author{Eric Thrane}
\affiliation{Monash Centre for Astrophysics, School of Physics and Astronomy, Monash University, VIC 3800, Australia}
\affiliation{OzGrav: The ARC Centre of Excellence for Gravitational-Wave Discovery, Clayton, VIC 3800, Australia}

\begin{abstract}
In order to extract information about the properties of compact binaries, we must estimate the noise power spectral density of gravitational-wave data, which depends on the properties of the gravitational-wave detector.
In practice, it is not possible to know this perfectly, only to estimate it from the data.
Multiple estimation methods are commonly used and each has a corresponding statistical uncertainty.
However, this uncertainty is widely ignored when measuring the physical parameters describing compact binary coalescences, and the appropriate likelihoods which account for the uncertainty are not well known.
In order to perform increasingly precise astrophysical inference and model selection, it will be essential to account for this uncertainty.
In this work, we derive the correct likelihood for one of the most widely used estimation methods in gravitational-wave transient analysis, the median average.
We demonstrate that simulated Gaussian noise follows the predicted distributions.
We then examine real gravitational-wave data at and around the time of GW151012, a relatively low-significance binary black hole merger event.
We show that the data are well described by stationary-Gaussian noise and explore the impact of different noise power spectral density estimation methods on the astrophysical inferences we draw about GW151012.
\end{abstract}

\maketitle

\section{Introduction}
The astrophysical parameters of compact binaries are inferred from gravitational-wave data using Bayesian inference.
A crucial first step for Bayesian inference is to choose the appropriate likelihood for our data.
In gravitational-wave transient data analysis this, typically, hinges on assumptions that the noise is Gaussian and stationary over the period being analysed~\citep{LIGOData}.
If these conditions are met, and if the noise power spectral density (PSD) were known exactly, the appropriate likelihood would be the Whittle likelihood~\citep{Romano2017}
\begin{equation}
    \mathcal{L}(\tilde{d}|\theta, P) = \frac{2}{\pi T P}\exp\left(-\frac{2|\tilde{d} - \tilde{\mu}(\theta)|^2}{TP}\right).
\label{eq:whittle}
\end{equation}
Here $P$ is the PSD, $\tilde{d}$ is the frequency domain interferometer data, $T$ is the duration of the data being analyzed, and $\tilde{\mu}$ is our model for the expected signal.

However, in practice, we do not have access to the true power spectral density of gravitational-wave detectors and so we have to rely on an empirical estimate.
There are two commonly used methods to compute these estimates.
The simplest method is to average over the power in neighbouring stretches of data to generate an ``off-source'' estimate.
This method assumes that the PSD does not vary over the duration being averaged and that there are no non-Gaussian features in the data.
The other commonly used method is to simultaneously fit the signal and power spectral density to obtain an ``on-source'' estimate, e.g.,~\citep{Littenberg2015}.
While this method does not involve analyzing as much data, and hence if less effected by non-stationarity and non-Gaussianity, it is far more computationally expensive.
In this work, we are going to focus on the former.

To generate an off-source PSD we typically either compute the mean or median average of neighbouring segments.
Taking the mean of neighbouring segments is a commonly used method (sometimes referred to as the Welch or Blackwell method~\citep{Welch1967}) in gravitational-wave data analysis and many other signal processing applications.
However, it is not widely used in gravitational-wave transient data analysis due to it's sensitivity to non-Gaussian transients, ``glitches,'' in the detector noise.
To mitigate the effect of these glitches a median average is instead used to compute the PSD as the median is more robust to the presence of large outliers.
However, there may be effective methods to either remove or exclude these non-Gaussian features~\cite{O1Stochastic, Usman2016, Pankow2018, Driggers2019, Davis2019, Sachdev2019, Venumadhav2019}.

The other assumption underlying Equation~\ref{eq:whittle} is that the PSD does not change over time; in other words, the data are stationary.
In practice, the PSD of real interferometers varies over the time scale of minutes, and so care must be taken when estimating the PSD using longer stretches of time~\cite{Littenberg2015, LIGOData, Vajente2020}.
Methods for mitigating this non-stationarity have also been considered previously~\cite{Venumadhav2019}.
In this paper, we ignore these possible effects and assume the data we look at are both Gaussian and stationary.

In~\citeauthor{Chatziioannou2019}, the authors compare Advanced LIGO data whitened with a median estimated off-source PSD with on-source PSD estimation methods.
They show that the data whitened using the median PSD does not follow a unit normal distribution.
They argue that this difference is due to non-Gaussianity and non-stationarity in the data.
However, data whitened using an off-source are known to follow a non-Normal distribution \textit{even for Gaussian noise}.
For a mean average, the whitened Gaussian-noise data follow a Student's $t$-distribution, and the correct likelihood to use is the Student-Rayleigh distribution~\citep{Rover2011a, Rover2011b, Yamamoto2016, Banagiri2019}.
In this work, we demonstrate that data whitened with a median PSD estimate follows a different distribution, and we show how to marginalise over the uncertainty in this estimated PSD to obtain the correct likelihood for stationary, Gaussian noise.

On-source PSD estimation using the {\tt BayesLine} algorithm~\citep{Littenberg2015} marginalizes over a prior which models the PSD as a combination of a slowly varying spline and Lorentzians to fit sharp spectral features.
This algorithm is generally combined with the {\tt BayesWave} algorithm~\citep{Cornish2015} to fit astrophysical and terrestrial transients simultaneously with the PSD.
However, this does not allow direct inference of physical parameters describing compact binary coalescences, e.g., the masses and spins of merging black holes.

A common approach in compact binary data analysis is to take a median average of the posterior distribution for the PSD obtained using {\tt BayesLine} rather than an off-source averaged PSD, e.g.~\citep{GWTC1}.
Recently,~\citeauthor{Biscoveanu2020} introduced a method to marginalize over the uncertainty in these on-source PSDs estimates.
However, this is done at significant computational cost requiring $\sim 200\times$ the computational resources as a standard analysis.
Additionally, under the formalism presented there, it is not possible to compute the Bayesian evidences necessary to perform model comparison.

The remainder of this paper is structured as follows.
In Section~\ref{sec:derivations}, we derive the appropriate distributions for the likelihood and whitened data after marginalising over the uncertainty in a median (and/or mean) PSD estimate.
We provide a brief introduction to Bayesian inference in the context of gravitational-wave astronomy in Section~\ref{sec:inference}.
We then demonstrate the efficacy of our formalism by applying it to simulated Gaussian data in Section~\ref{sec:gaussian-demonstration}.
Following this, in Section~\ref{sec:151012} we consider a case study using real Advanced LIGO data.
We analyze the marginal gravitational-wave candidate GW151012 with both mean and median PSD estimates to understand the effect of marginalizing over the statistical uncertainty and of using the different estimation techniques.
This event is convenient for our present purposes since the effects we seek to study are most prominent for marginal signals like GW151012.
Some closing comments are then provided in Section~\ref{sec:discussion}.

\section{Formalism}\label{sec:derivations}

\subsection{Gaussian noise}

For stationary Gaussian noise $n(t)$, if we do not manipulate the data in any way before performing a Fourier transform, the noise covariance can be written in the frequency domain as
\begin{equation}
    C(f, f') = \left< \tilde{n}(f) \tilde{n}^{*}(f') \right> = P(f) \delta(f - f').
\end{equation}
The angle braces denote an ensemble average over realizations.
In practice, we work with discrete Fourier transforms and noise covariance matrices
\begin{equation}
    C_{ij} = \left< \tilde{n}_i \tilde{n}^{*}_j \right> = \frac{T}{4} P_i \delta_{ij} = \frac{T}{4} A_{i} \delta_{ij} A_{j}
\end{equation}
Here $T$ is the duration over which the discrete Fourier transforms is performed, $i,j$ index frequency bins, and $A_i = P^{1/2}_{i}$ is the noise amplitude spectral density (ASD).

For real data, a number of manipulations are performed before the data are Fourier transformed, which makes things more complicated.
The data are band-passed and windowed in the time domain to prevent aliasing and spectral leakage~\cite{LIGOData}.
As long as the frequency limits of the band-pass filters do not overlap with the frequency range of interest, the band-passing can be ignored.
However, the window applied to the data must be considered.
Since the window is multiplicative in the time domain, there is a corresponding convolution in the frequency domain,
\begin{align}
    C^{w}_{ij} &= \left< (\tilde{n} * \tilde{W})^{*}_{i} (\tilde{n} * \tilde{W})_{j} \right> = \frac{T}{4} A_{i} \mathcal{T}_{ij} A_{j}, \\
     \mathcal{T}_{ij} &= \begin{cases}
        \tilde{W}_{i-j} & i\geq j\\
        \tilde{W}^{*}_{j - i} & i < j\\
    \end{cases}.
\end{align}
Where $\mathcal{T}$ is a Hermitian Toeplitz matrix and there is no implied summation over $i$ or $j$.
For a rectangular window $\mathcal{T}_{ij} = \delta_{ij}$ and the covariance matrix is diagonal.
For generic windows, there is a regular, predictable, off-diagonal power.
In reality, the effect of this is much smaller than the effects considered here, and inverting the covariance matrix is a significant computational challenge.
We leave a detailed analysis of the effect of non-rectangular windows on parameter estimation and model selection to a future study.

The real and imaginary components of the frequency-domain noise follow a normal distribution with variance matrix $P$,
\begin{equation}
    p(\tilde{n}_{i} | P_{i}) = \sqrt{\frac{2}{\pi T P_{i}}} \exp \left( - \frac{2 \tilde{n}^{2}_{i}}{T P_{i}} \right)
\label{eq:normal}
\end{equation}
This is {\emph not} the likelihood which we use when analyzing gravitational-wave transients as we need to simultaneously consider the real and imaginary components of the noise.
The likelihood is given by
\begin{equation}
    \mathcal{L}(\tilde{n}_{i}, \tilde{n}_{j} | C_{ij}) = \frac{2}{\pi T \det(C_{ij})} \exp \left( - \frac{1}{2} \tilde{n}_{i} C^{-1}_{ij} \tilde{n}^{*}_{j} \right).
\end{equation}
Since we assume the covariance matrix is diagonal this is often written in the simplified form known as the Whittle likelihood,
\begin{equation}
    \mathcal{L}(\tilde{n}_{i} | C_{i}) = \frac{2}{\pi T P_{i}} \exp \left( - \frac{2 |\tilde{n}_{i}|^{2}}{T P_i} \right).
\end{equation}
We note that this likelihood is normalized over the complex plane.
It is convenient to reduce to one dimension for visualisation purposes, so we note that the power of the noise $\mathcal{P}_{i} = |\tilde{n}_{i}|^2$ follows an exponential distribution,
\begin{equation}
    p(\mathcal{P}_{i} | P_{i}) = \frac{2}{TP_{i}} \exp \left( - \frac{2 \mathcal{P}_{i}}{T P_{i}} \right).
\end{equation}

All of the expressions above assume that there are no non-Gaussian signals in the data.
In order to include signals we simply make the substitution $n = d - \mu$ where $d$ is the data and $\mu$ is the signal.

Time-domain windows affect the noise and signal components differently.
We assume that the window is always applied such that the window does not cause any loss of signal power in the observing frequency band.
In addition to the correlation between different frequency bins induced by the window, there is a net power loss in the Gaussian noise given by the mean square value of the window function.
Care must be taken to consistently correct for this power loss to avoid biasing our inference, e.g.,~\cite{Talbot2020}.

Now that we have established which distributions we want to use when the PSD is known, we can address the distributions we want to use when the PSD is uncertain.

\subsection{Median PSD estimate}

The generic expression for the likelihood marginalised over uncertainty in an estimated PSD, $\hat{P}$, is
\begin{align}
    {\cal L}_P(\tilde{d}|\theta, \hat{P})
    &= \int_0^\infty dP \, {\cal L}(\tilde{d}|\theta, P) \pi(P|\hat{P}).
\label{eq:start}
\end{align}
Where ${\cal L}(\tilde{d}|\theta, P)$ is the likelihood of obtaining the data given model parameters $\theta$ and the true PSD $P$, as defined in Eq.~\ref{eq:whittle}, and $\pi(P|\hat{P})$ is our prior on the true PSD given the estimated PSD.
Similarly, using Equation~\ref{eq:normal}, we can write down an expression for the expected distribution of whitened strain residuals, $\tilde{\nu} = \tilde{n} / \hat{P}^{1/2}$,
\begin{align}
    p_{P}\left(\tilde{\nu} | \hat{P}\right)
    &= \int_0^\infty dP \, p(\tilde{n}|P) \pi(P|\hat{P}).
\label{eq:start-strain}
\end{align}
Here $\tilde{n}$ is the frequency-domain data after removing any signals present.

First we need to define the estimated PSD
\begin{align}
\hat{P} = \frac{\text{median}(P_\ell)}{\alpha}.
\end{align}
Where 
\begin{equation}
    \alpha = \sum^{N}_{\ell=1} \frac{(-1)^{\ell}}{\ell}
\end{equation}
is a factor to account for the median being a biased estimator of the mean (see, e.g., Appendix B of~\cite{Allen2012}), and $\ell$ indexes the segments being averaged over.
For simplicity, we assume that we are computing the median of an odd number, $N$, of non-overlapping stretches ensuring $\alpha > 0$.

It is convenient to work with a regularised version of the PSD, 
\begin{equation}
Q = 2 \hat{P} / P.
\end{equation}
Since the data are assumed to follow a zero-mean Gaussian distribution with variance $P$, the $Q_{i}$ are drawn from a $\chi^2$ distribution of order 2,
\begin{align}
    p(Q) = \chi^2_{2}\left(Q\right) = \frac{1}{2} \exp\left(-\frac{Q}{2}\right).
\end{align}
Additionally, we define the usual cumulative distribution function, $\Phi$, and  survival function, $S$, for this quantity,
\begin{align}
    \Phi(Q) &= \int_{0}^{Q} d{Q'} p(Q') = 1 - \exp\left(-\frac{Q}{2}\right) \\
    S(Q) &= \int_{Q}^{\infty} d{Q'} p(Q') = \exp\left(-\frac{Q}{2}\right).
\end{align}

The probability of the median of an odd number of segments follows the median order statistic.
This is the probability of the getting the median value from the distribution multiplied by the probability of having $m = (N - 1) / 2$ measurements less than $\hat{Q}$ and $m$ measurements larger than $\hat{Q}$.
Symbolically, this is
\begin{align}
    \pi(Q|\hat{P})
    &= \frac{p(Q)}{2 \hat{P}} \frac{\Phi(Q)^{m} S(Q)^m}{\Beta(m+1, m+1)}\\
    \label{eq:pqmed}
    &= \frac{\left(1 - e^{-\frac{1}{2} Q}\right)^{m} e^{-\frac{(m+1)}{2}Q}}{4 \hat{P} \Beta(m+1, m+1)}\\
    &= \sum_{k=0}^{m} \binom{m}{k} \frac{(-1)^{k} e^{-\frac{(m+k+1)}{2}Q}}{4 \hat{P} \Beta(m+1, m+1)} .
\end{align}
Where $\Beta$ is the Beta function and in the last line we perform a binomial expansion.
The final piece we need is to relate our prior on $Q$ to our prior on $P$,
\begin{equation}
    \pi(P|\hat{P}) dP = \pi(Q | \hat{P}) dQ.
\end{equation}

Substituting this expression into Equation~\ref{eq:start}, the PSD-marginalized likelihood is
\begin{align}
    {\cal L}_P
    &= \int_0^\infty dQ \,
    \frac{Q}{8\pi \hat{P}}
    \frac{\left(1 - e^{-\frac{Q}{2}}\right)^{m} e^{- \frac{Q}{2} \left(\frac{1}{2}\left|\tilde{\nu}\right|^2 + m + 1\right)}}{\Beta(m+1, m+1)} \label{eq:numerical-median-marginalised}\\
    &= \sum_{k=0}^{m} \binom{m}{k} \frac{(-1)^{k}}{2\pi\hat{P}}
    \frac{\left(m + k + 1 + \frac{\left|\tilde{\nu}\right|^2}{2} \right)^{-2}}{\Beta(m+1, m+1)}, \label{eq:analytic-median-marginalised}
\end{align}
and using Equation~\ref{eq:start-strain}, the distribution of whitened residuals is
\begin{align}
    p(\tilde{n})
    &= \int_0^\infty dQ \,
    \frac{1}{4}\sqrt{\frac{Q}{2 \pi \hat{P}}}
    \frac{\left(1 - e^{-\frac{Q}{2}}\right)^{m} e^{- \frac{Q}{2} \left(\frac{1}{2}\left|\tilde{\nu}\right|^2 + m + 1\right)}}{\Beta(m+1, m+1)} \label{eq:numerical-median-marginalised-whiten}\\
    &= \sum_{k=0}^{m} \binom{m}{k} \frac{(-1)^{k}}{\sqrt{2\pi\hat{P}}}
    \frac{\left(m + k + 1 + \frac{\left|\tilde{\nu}\right|^2}{2} \right)^{-3/2}}{\Beta(m+1, m+1)} \label{eq:analytic-median-marginalised-whiten}.
\end{align}

While the final expressions Equations~\ref{eq:analytic-median-marginalised} and~\ref{eq:analytic-median-marginalised-whiten} are exact closed form solutions, they are numerically unstable and cannot be safely computed for $m \gtrsim 15$.
We therefore simply construct an interpolant over numerically computed values of the integrals in Equations~\ref{eq:numerical-median-marginalised} and \ref{eq:numerical-median-marginalised-whiten}, which can be rapidly evaluated at run time.

\subsection{Mean PSD estimate}

The appropriate distribution to use for a mean averaged PSD has been discussed and independently derived multiple times in the literature, e.g.,~\cite{Rover2011b, Banagiri2019}; in this work, we just quote the relevant results.
For a mean estimate:
\begin{equation}
    \hat{P} = \frac{1}{N}\sum^{N}_{i=1} P_{i}.
\end{equation}
\begin{equation}
    \pi(P|\hat{P}) dP = \pi(Q|\hat{P}) dQ = \chi^2_{2N}(Q) dQ.
\end{equation}
\begin{align}
    {\cal L}_P = \frac{1}{2 \pi \hat{P}} 
    \left(1 + \frac{\left|\tilde{\nu}\right|^2}{2N}\right)^{-(N + 1)},
\label{eq:student-rayleigh}
\end{align}
\begin{align}
    p(\tilde{\nu}) = \frac{\Gamma(N + 1/2)}{\sqrt{2 \pi N  \hat{P}}\Gamma(N)}
    \left(1 + \frac{\left|\tilde{\nu}\right|^2}{2N}\right)^{-(N + 1/2)}.
\label{eq:student-t}
\end{align}

Equation~\ref{eq:student-rayleigh} is the Student-Rayleigh distribution and Equation~\ref{eq:student-t} is the Student's $t$-distribution with $2N$ degrees of freedom, two degrees for each segment being averaged over, coming from the real and imaginary components of the frequency domain strain.

\subsection{Limiting cases}

When $N=1$ the mean and median are the same and so Equation~\ref{eq:analytic-median-marginalised-whiten} should reduce to a Student's $t$-distribution with two degrees of freedom.
As expected, we find
\begin{align}
    p(\tilde{n})
    &= \frac{1}{\sqrt{2\pi\hat{P}}} \left(1 + \frac{|\tilde{\nu}|^2}{2}\right)^{-3/2}.
\end{align}
Analogously, the PSD uncertainty marginalised likelihoods also match and are both
\begin{align}
    {\cal L}_P = \frac{1}{2 \pi  \hat{P}} 
    \left(1 + \frac{\left|\tilde{\nu}\right|^2}{2}\right)^{-2}.
\end{align}

The other important limiting case is when $N \rightarrow \infty$.
It is a well-known result that the Student's t-distribution converges to a Gaussian in this limit; this follows from Taylor-expanding the distribution.
Performing a similar expansion, it is possible to demonstrate that the Student-Rayleigh distribution converges to the Whittle likelihood.
We numerically confirm that (\ref{eq:analytic-median-marginalised-whiten}) and (\ref{eq:analytic-median-marginalised}) also converge to a Gaussian distribution and Whittle likelihood respectively.

\section{Bayesian inference for gravitational-wave transients}~\label{sec:inference}

In the previous section, we derived likelihood functions, marginalized over the statistical uncertainty in an estimate of the PSD.
These likelihood functions $\mathcal{L}(\tilde{d}|\theta, \hat{P}, \mathcal{M})$ are the probability of obtaining data $\tilde{d}$ given a signal model $\mathcal{M}$ described by parameters $\theta$ and a PSD estimate $\hat{P}$.
However, we are generally interested in measuring the source-model parameters and performing model comparison.
Using Bayes' theorem we get
\begin{equation}
    p(\theta|\tilde{d}, \hat{P}, \mathcal{M}) = \frac{\mathcal{L}(\tilde{d}|\theta, \hat{P}, \mathcal{M}) \pi(\theta, \hat{P}, \mathcal{M})}{\mathcal{Z}(\tilde{d} | \hat{P}, \mathcal{M})}.
\end{equation}
The term on the LHS, $p(\theta|\tilde{d}, \hat{P}, \mathcal{M})$, is the posterior probability distribution, the probability of the parameters describing the model given the data.
The term $\pi(\theta, \hat{P}, \mathcal{M})$ is our prior distribution which is based on our expectation before analysing the data.
The term $\mathcal{Z}(\tilde{d} | \hat{P}, \mathcal{M})$ is the evidence for the data given the model $\mathcal{M}$.

The evidence is used for model comparison by computing Bayes factors for two models
\begin{equation}
    BF^{1}_{0} = \frac{\mathcal{Z}(\tilde{d} | \hat{P}, \mathcal{M}_{1})}{\mathcal{Z}(\tilde{d} | \hat{P}, \mathcal{M}_{0})}.
\end{equation}
While the Bayes factor is often used for model selection, strictly speaking, we should compare the probability of the model given the data, rather than the probability of the data given the model.
This is given by the odds
\begin{equation}
    \mathcal{O}^{1}_{0} = \frac{\mathcal{Z}(\tilde{d} | \hat{P}, \mathcal{M}_{1}) }{\mathcal{Z}(\tilde{d} | \hat{P}, \mathcal{M}_{0})} \frac{\pi(\mathcal{M}_{1} | \hat{P})}{\pi(\mathcal{M}_{0} | \hat{P})},
\end{equation}
which is the Bayes factor comparing the two models multiplied by the prior odds.
Throughout this work, we will assume all models have equal prior odds and so the odds reduces to the Bayes factor.

Finally, we define the coherent vs incoherent Bayes factor~\cite{Veitch2015}, BCI, as a measure of the relative probability that the data contain a coherent signal or incoherent signals in different detectors,
\begin{equation}
    {\rm BCI} = \frac{\mathcal{Z}(\{\tilde{d}_k\} | \{\hat{P}_k\}, \mathcal{M}))}{\Pi_k \mathcal{Z}(\tilde{d}_k | \hat{P}_k, \mathcal{M}))}.
\end{equation}
Here, the $k$ index multiple independent interferometers, e.g., LIGO Hanford and LIGO Livingston.
As in~\cite{Isi2018} we assume that any incoherent signals are described by the same model as the coherent signals, however, this is not necessarily the case~\cite{Ashton2019b}.
We note that the BCI is not used as the final discriminator between the coherent and incoherent models as it is missing a prior for the relative rates of coherent and incoherent signals.
In~\cite{Isi2018, Ashton2019b} the priors on rate are empirically calibrated delta functions, however, in~\cite{Smith2018} the authors fit the rates of coherent and incoherent signals. 

In general relativity, non-eccentric binary black hole coalescences are fully described by fifteen parameters.
Eight parameters which describe the ``intrinsic'' properties of the binary (two masses and two three-dimensional angular momentum vectors), and seven ``extrinsic'' parameters to specify the position, orientation, and coalescence time of the binary relative to Earth.
This parameter space is typically explored using stochastic samplers using either Markov-chain Monte Carlo~\cite{Hastings1970} or nested sampling~\cite{Skilling2006}.

In order to improve the convergence of the sampling and accelerate our inference, it is possible to use a modification of the Whittle likelihood which is marginalized over the coalescence time, orbital phase, and distance of the source~\cite{Veitch2015, Thrane2019}.
It is not possible to perform these marginalizations as easily while also marginalizing over uncertainty in the PSD.
Therefore, in this work, we perform our inference in two stages following~\cite{Payne2019}.
\begin{enumerate}
    \item First, we analyze the data using the Whittle likelihood marginalized over coalescence time, binary orbital phase, and distance to obtain samples from the posterior distribution and an estimate of the signal evidence and Gaussian noise evidence.
Posterior distributions for these marginalized parameters are then recomputed in post-processing.
We use {\tt dynesty}~\cite{Speagle2019}, an implementation of the nested sampling algorithm, as implemented in {\tt Bilby}~\cite{Ashton2019a} to sample the space.
    \item After this, we importance resample the posterior obtained in the previous step by the ratio of the PSD-marginalised likelihood to the Whittle likelihood to obtain posterior samples and an evidence which include the marginalisation over the statistical uncertainty.
\end{enumerate}

We note that the importance sampling in step 2 only works when resampling to a distribution which is similar to the original posterior distribution.
We quantify this by evaluating the efficiency of the resampling and the number of effective samples from the PSD marginalised posterior.
Since the marginalized likelihoods converge to the non-marginalized likelihood when averaging many segments, we expect the resampling to be efficient.
This method also generically gives a much smaller uncertainty on the Bayes factor comparing the two models than would be obtained by performing two independent sampling runs~\cite{Huebner2020}.
A similar method has previously been employed for cosmological inference in~\cite{Sellentin2016} to marginalize over uncertainty in an estimated covariance matrix.

\section{Demonstration with Gaussian noise}\label{sec:gaussian-demonstration}

\begin{figure}
    \includegraphics[width=0.98\linewidth]{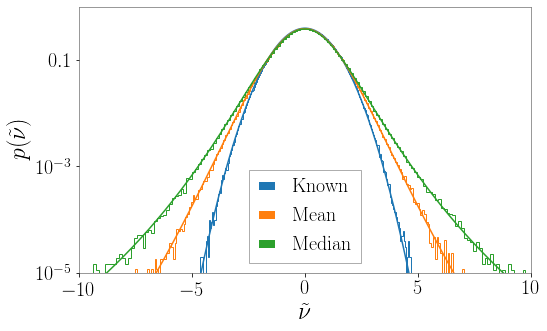}
    \includegraphics[width=0.98\linewidth]{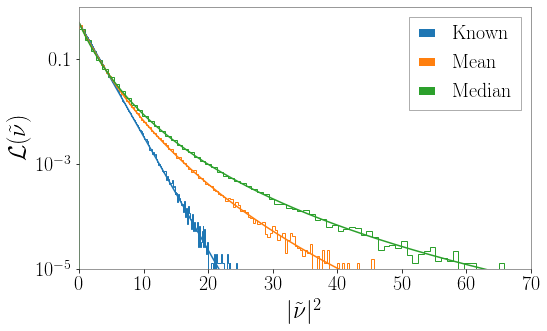}
    \includegraphics[width=0.98\linewidth]{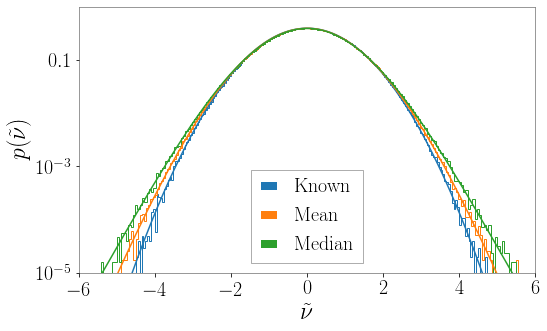}
    \includegraphics[width=0.98\linewidth]{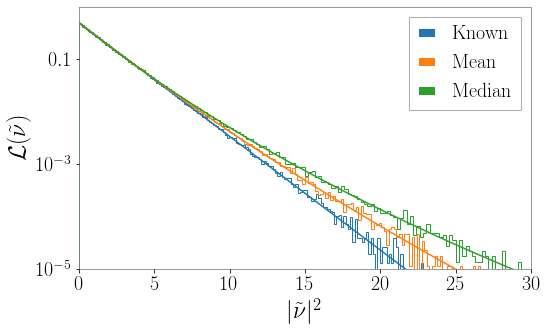}
    \caption{
    Comparison of the expected and empirical distribution of whitened frequency-domain strain (first and third panels) and whitened noise power (second and fourth panels) when using three different PSDs for whitening and simulated Gaussian noise colored to the known PSD.
    The three PSDs used are the true ``known'' PSD (blue), a mean estimate (orange), and a median estimate (green).
    The number of averages used to generate the mean and median estimates are 7 (top pair), and 31 (bottom pair).
    In all cases the data follow the predicted distributions.
    }
    \label{fig:data-whitening}
\end{figure}

\begin{figure*}
    \includegraphics[width=0.48\linewidth]{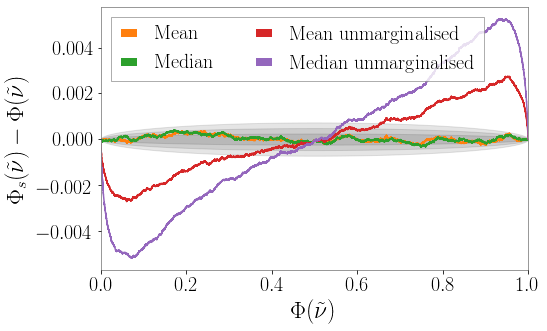}
    \includegraphics[width=0.48\linewidth]{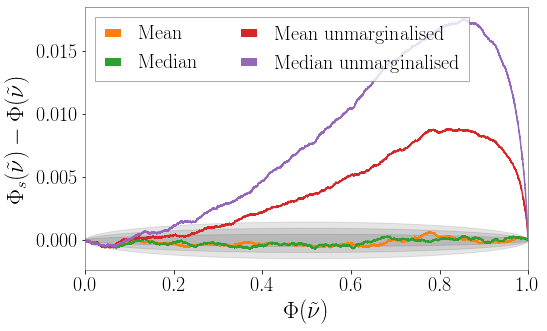}
    \includegraphics[width=0.48\linewidth]{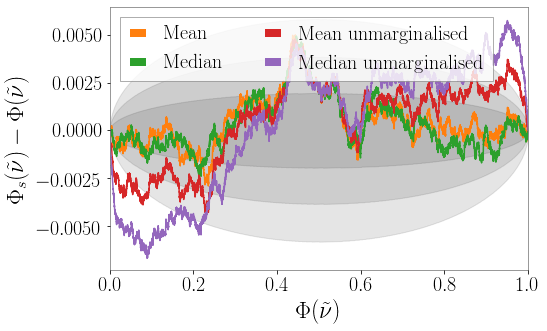}
    \includegraphics[width=0.48\linewidth]{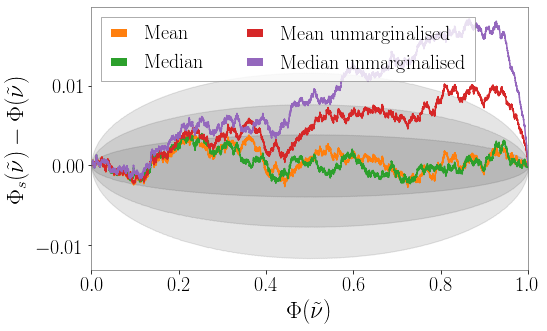}
    \caption{
    Difference between the empirical and expected cumulative distributions of whitened frequency-domain strain (left) and whitened noise power (right) when using two different PSDs for whitening and simulated Gaussian noise colored to a known PSD.
    In the top panels we consider 512 seconds of simulated data and in the bottom panels we consider 8 seconds of simulated data.
    The two PSDs used are a mean estimate (orange/red), and a median estimate (green/purple).
    The gray shaded regions show the expected $1\sigma$, $2\sigma$, and $3\sigma$ uncertainties.
    We average over 31 realisations to generate the PSDs.
    There are significant deviations when not marginalizing over the uncertainty in the PSD once enough data are considered.
    }
    \label{fig:gaussian-ppp}
\end{figure*}

\begin{figure}
    \includegraphics[width=\linewidth]{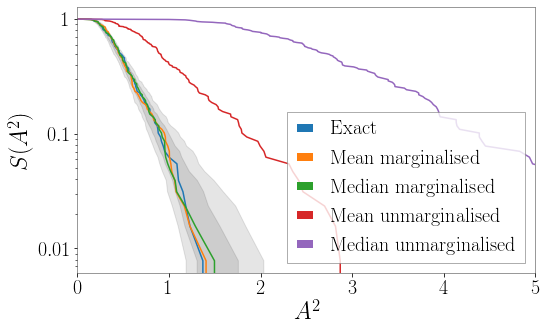}
    \includegraphics[width=\linewidth]{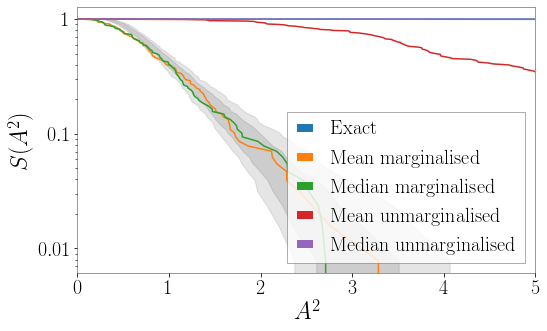}
    \caption{
    The survival function of the Anderson-Darling statistic comparing Gaussian noise whitened with three PSDs with the distributions which do/do not marginalize over the statistical uncertainty in the PSD estimate.
    In the top panel, we analyze the real and imaginary components of the whitened strain.
    In the bottom panel, we analyze whitened power.
    The blue curve is generated using data whitened by the exact known PSD.
    The orange/green curves are generated by comparing data whitened by the mean/median PSD estimates with the distributions which marginalize over the statistical uncertainty.
    The red/purple curves are generated by comparing data whitened by the mean/median PSD estimates with the distributions which do not marginalize over the statistical uncertainty.
    The grey shaded regions show the expected $1\sigma$ and $2\sigma$ uncertainties.
    We average over 31 realisations to generate the PSDs.
    The Anderson-Darling statistic does not follow the expected distribution when not marginalizing over the uncertainty in the PSD.
    }
    \label{fig:anderson-darling}
\end{figure}

To demonstrate the accuracy of the methods described in Section~\ref{sec:derivations} we analyze simulated Gaussian noise colored by the Advanced LIGO design sensitivity PSD~\cite{ObservingScenarios}.
Following~\citep{Chatziioannou2019}, we perform three tests on data whitened using median and mean PSD estimates for verification.
As an extension to the analysis presented in~\citep{Chatziioannou2019}, we consider the whitened power, $|\tilde{\nu}|^2$ in addition to the real and imaginary components of the whitened strain $\tilde{\nu}$.
For all estimated PSDs, we average non-overlapping segments with the same duration as the analysis segment.

First, we perform a visual test of the whitened data.
In Figure~\ref{fig:data-whitening}, we show the distribution of the real and imaginary components of the whitened strain (first and third panels) and whitened power (second and fourth panels) along with the theoretical expectations.
In the top (bottom) pair of panels, we average over 7 (31) independent noise realisations.
We see that the data whitened using the off-source estimates follow the expected distributions in each case.

In Figure~\ref{fig:gaussian-ppp} we show the difference between the empirical and expected cumulative distribution functions, $\Phi_{s} - \Phi$, plotted against the expected cumulative distribution function for the same data as in Figure~\ref{fig:data-whitening}.
In orange and red we compare the data whitened with the mean PSD estimate with the expected distributions with and without marginalizing over the uncertainty in the PSD respectively.
In green and purple we compare the data whitened with the median PSD estimate with the expected distributions with and without marginalizing over the uncertainty in the PSD respectively.
The grey regions indicate the expected $1\sigma$, $2\sigma$, and $3\sigma$ fluctuations.
For both PSD estimation methods, we see that the data agree better with the distributions which marginalize over the uncertainty in the PSD.

When comparing the marginalized distributions to the non-marginalized distributions we see two clear deviations from the expected behaviour.
The whitened strain uncertainty-marginalized distribution has wider, symmetric, tails than a normal distribution leading to the negative $\Phi_{s} - \Phi$ for small $\tilde{\nu}$ and positive $\Phi_{s} - \Phi$ for large $\tilde{\nu}$.
The distribution of the whitened power, however, only has a wide tail out to large $|\tilde{\nu}|$, leading to the positive $\Phi_{s} - \Phi$ for large $\tilde{\nu}$.

To quantify the similarity of the data to the expected distributions we compute the Anderson-Darling statistic
\begin{equation}
    A^2 = N \int^{\infty}_{-\infty} d\Phi \, \frac{(\Phi_{s} - \Phi)^2}{\Phi(1 - \Phi)}.
\end{equation}
Here $N$ is the number of samples, in this case, the number of frequency bins.
The numerator is the square of the quantity on the vertical axis of Figure~\ref{fig:gaussian-ppp} and the integral is over the horizontal axis.

In Figure~\ref{fig:anderson-darling} we show the survival function of the distribution of the Anderson-Darling statistic for four cases: for both the mean and median PSD estimation methods we compare the distribution of the whitened strain to a unit normal distribution and the expected distribution as described in the previous section for the whitened strain (left) and whitened power (right).
We also show the expected distribution if the two distributions are the same.

We note that the gradient of the expected distribution of the Anderson-Darling statistic is steeper for the whitened strain than for the whitened power.
When applying a window the data before performing the discrete Fourier transforms, the real and imaginary components of the frequency domain strain are no longer independent, reducing the appropriate value of $N$ by a factor of two (see, Appendix A of~\cite{Talbot2020}).
We, therefore, avoid the case identified in~\cite{Chatziioannou2019} where the distribution of $\nu$ appeared to match the correct distribution better than possible.

\section{A case study - GW151012}\label{sec:151012}

To examine the effect of non-Gaussianity and non-stationarity on the noise properties of real gravitational-wave detectors, we analyze the data in the two Advanced LIGO interferometers~\cite{AdvancedLIGO} at and around the time of GW151012~\cite{O1BBH}, the lowest significance binary black hole merger included in the first gravitational-wave transient catalog~\cite{GWTC1}.
We analyze $\unit[256]{s}$ of data ending $\unit[2]{s}$ after the merger time.
We subdivide the data into 32 $\unit[8]{s}$ chunks, the first 31 chunks are used to compute the PSD and the final $\unit[8]{s}$ are the on-source data.

We apply a Tukey window with a roll off of $\unit[0.2]{s}$ to each of the chunks to suppress spectral leakage.
We then fast-Fourier transform the windowed time-domain strain before averaging the PSD chunks.
We do not apply the conventional window amplitude correction factor to either the PSD or the data.
After applying the fast-Fourier transform, we remove all data below $\unit[20]{Hz}$ and above $\unit[1024]{Hz}$.

\begin{figure}[htbp]
    \centering
    \includegraphics[width=\linewidth]{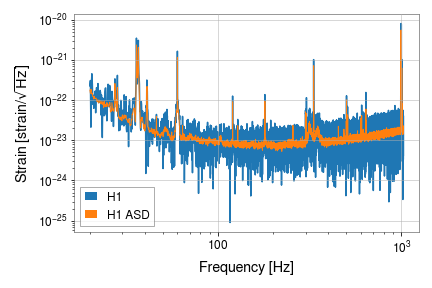}
    \includegraphics[width=\linewidth]{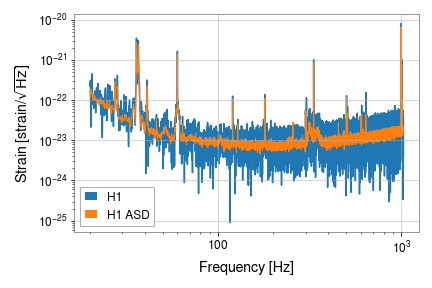}
    \caption{
    Frequency domain strain power (blue) and mean (top) and median (bottom) estimated power spectral densities (orange) for LIGO Hanford at the time of GW151012.
    }
    \label{fig:151012-H1-psds}
\end{figure}

\begin{figure}[htbp]
    \centering
    \includegraphics[width=\linewidth]{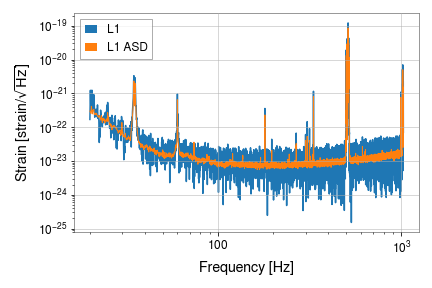}
    \includegraphics[width=\linewidth]{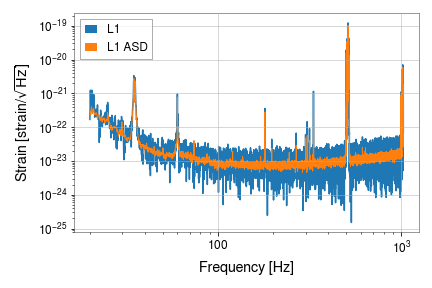}
    \caption{
    Frequency domain strain power (blue) and mean (top) and median (bottom) estimated power spectral densities (orange) for LIGO Livingston at the time of GW151012.
    }
    \label{fig:151012-L1-psds}
\end{figure}

The resulting PSDs and the power in the on-source data are shown in Figures~\ref{fig:151012-H1-psds} and~\ref{fig:151012-L1-psds}.
Figure~\ref{fig:151012-H1-psds} shows data from the LIGO Hanford interferometer and Figure~\ref{fig:151012-L1-psds} data from the LIGO Livingston interferometer.
The orange curves show the mean estimated PSDs and the green show the median estimated PSDs.
All the PSD estimates are at the centre of the scatter in the on-source data, as expected.
We note that the width of the scatter on the mean PSDs is slightly smaller than for the median due to the slower convergence of the median estimate.

\subsection{Data quality tests}

The main reason for using a median estimate over a mean estimate is to mitigate the effect of large non-Gaussian transients.
However, the formalism derived above is invalid if there is a large outlier in the data being averaged over.
Therefore, we try to identify if any of the segments are clear outliers.
We compute the power per segment divided by the mean power in all the other segments.
This is essentially testing how well the data in each of the segments is whitened by the data in the other segments.
We apply an empirically tuned threshold of $1.5$ for the mean whitened power per segment.
Any segment with a mean power above this value we discard and repeat the test.
We identify that one segment of the Hanford data which fails this test with a mean whitened power of $2.66$.
Visual inspection reveals that this segment has a larger amplitude than all the others below $\sim 100$Hz.
No significant outliers are present in the selected Livingston data.

Additional tests of the quality are possible and performed routinely during gravitational-wave data analysis.
For example, researchers often remove specific frequency bins if the noise at that frequency is known to be non-Gaussian, e.g., around the frequency (and higher harmonics) of mains electricity~\cite{LIGOData}.
A possible extension would be to use the normalised average power used above to track non-stationarity in the data, a similar method is used in in~\cite{Venumadhav2019}.
Implementing further data quality cuts and vetoes will improve the quality of our off-source PSD estimates and is an interesting avenue for further study.

\subsection{Data whitening}

\begin{figure*}[htbp]
    \centering
    \includegraphics[width=\linewidth]{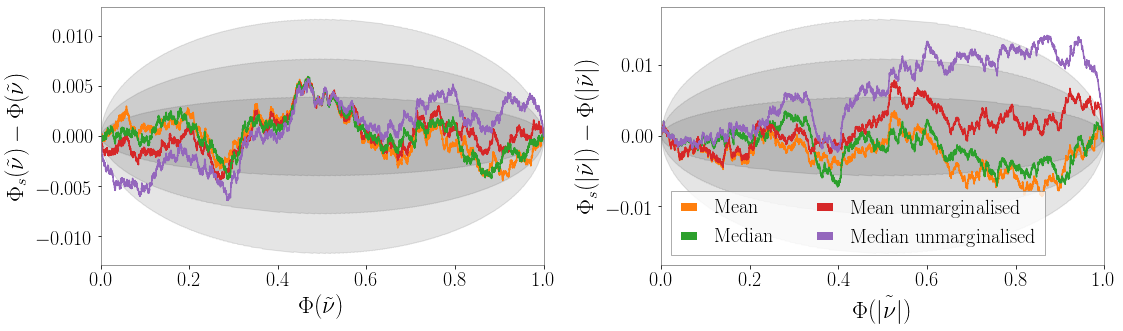}
    \includegraphics[width=\linewidth]{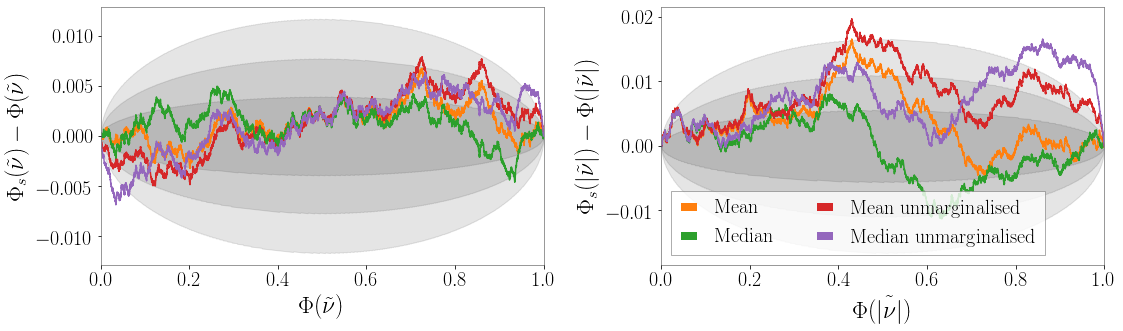}
    \caption{
    The difference between the empirical and analytic cumulative distribution functions of whitened strain $\tilde{\nu}$ and whitened power $|\tilde{\nu}|^2$ in the LIGO Hanford (top) and Livingston (bottom) interferometers with two different power spectral density estimation methods at the time of GW151012.
    The gray shaded regions show the expected $1\sigma$, $2\sigma$, and $3\sigma$ fluctuations.
    For the orange and green curves, the data are compared with the distributions which marginalise over uncertainty in the power spectral density estimate.
    For the red and purple curves, the data are compared with the distributions which do not marginalize over uncertainty in the power spectral density estimate.
    We note that the latter pair of curves deviate from the $3\sigma$ region, while the former does not.
    The data are well described by a stationary Gaussian process when marginalising over uncertainty in the power spectral density.
    }
    \label{fig:151012-data}
\end{figure*}

\begin{table*}[]
    \centering
    \begin{tabular}{c|c|c|c|c}
         & Mean vs marginalised & Median vs marginalised & Mean vs not marginalised & Median vs not marginalised \\ \hline\hline
         Livingston strain & 0.53 & 0.46 & 1.62 & 2.66 \\
         Hanford strain & 0.56 & 0.42 & 0.48 & 2.21 \\ \hline
         Livingston power & 1.48 & 1.03 & 4.22 & 6.71 \\
         Hanford power & 0.79 & 0.48 & 0.84 & 5.55
    \end{tabular}
    \caption{
    Values of the Anderson-Darling statistic for the whitened strain for the mean and median marginalised likelihood at the time of GW151012.
    Larger values of the Anderson-Darling statistic indicate comparatively worse agreement.
    The marginalized distributions match the data better.
    }
    \label{tab:anderson-darling-151012}
\end{table*}

We repeat the tests performed in Section~\ref{sec:gaussian-demonstration} on the data.
In Figure~\ref{fig:151012-data}, we show the deviations from the expected cumulative distribution functions for the data from the Hanford (top) and Livingston (bottom) interferometers.
On the left, we show the real and imaginary components of the whitened strain and on the right the whitened power.
In orange we show the difference between the empirical mean-estimated PSD whitened data and expected mean-marginalised distributions (\ref{eq:student-t}, \ref{eq:student-rayleigh}), in green the difference between the empirical median-estimated PSD whitened data and expected median-marginalised distributions (\ref{eq:analytic-median-marginalised-whiten}, \ref{eq:analytic-median-marginalised}).
In red (purple) we compare the data whitened using the mean- (median-) estimated PSDs with the distributions which do not account for the uncertainty.
In gray we show the $1\sigma$, $2\sigma$, and $3\sigma$ expected deviations.

In Table~\ref{tab:anderson-darling-151012}, we quote the corresponding values of the Anderson-Darling statistic for each of these lines.
We see that the largest deviations are observed when using data whitened with a median estimated PSD and compared to the non-marginalised distributions.
We also see that, with the exception of the strain components in Hanford, the Anderson-Darling statistic is always smaller when using the appropriate marginalised distributions.

\subsection{Impact on inference}

\begin{figure}
    \centering
    \includegraphics[width=\linewidth]{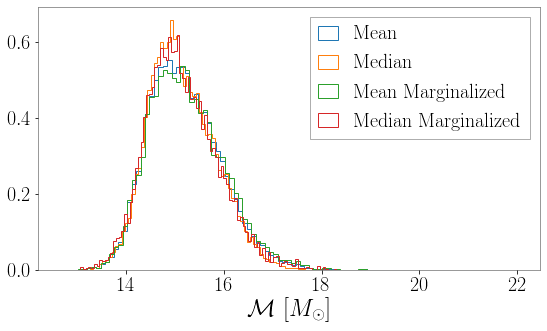}
    \includegraphics[width=\linewidth]{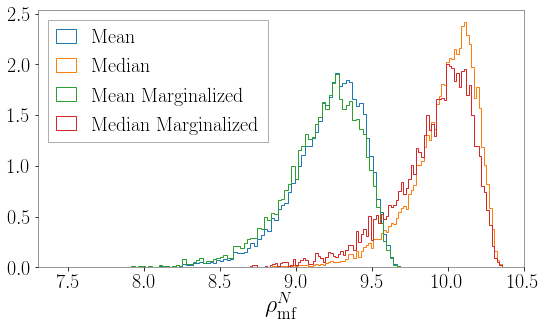}
    \caption{The posterior distribution for chirp mass (top) and matched filter SNR (bottom) for GW151012 for four different models.
    In blue and green we use the mean estimated PSD while in yellow and red we use the median estimate.
    In blue and yellow we neglect the uncertainty in the PSD estimate and in green and red we marginalise over the appropriate statistical uncertainty.
    We note that in both cases, marginalising over the uncertainty increases the width of the chirp mass posterior and decreases the average SNR.
    The mean estimated PSD gives a wider chirp mass posterior than the median PSD estimate and we see a corresponding decrease in the recovered matched filter SNR.
    }
    \label{fig:151012_intrinsic}
\end{figure}

\begin{figure}
    \centering
    \includegraphics[width=\linewidth]{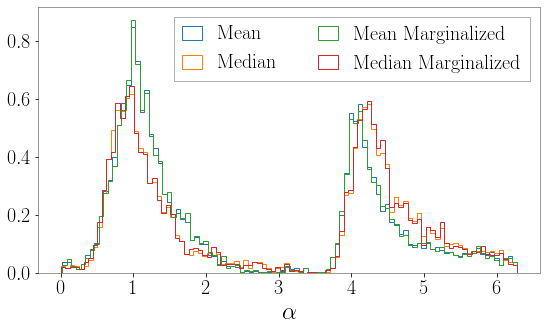}
    \includegraphics[width=\linewidth]{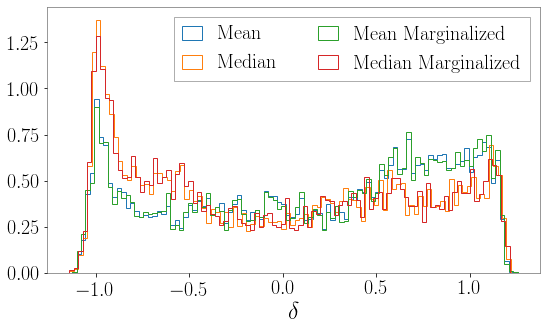}
    \caption{The posterior distribution for right ascension (top) and declination (bottom) for GW151012 for four different models.
    In blue and green we use the mean estimated PSD while in yellow and red we use the median estimate.
    In blue and yellow we neglect the uncertainty in the PSD estimate and in green and red we marginalise over the appropriate statistical uncertainty.
    In this case, marginalising over the uncertainty does not make a large difference to the inferred posterior distributions.
    However, the different PSD estimation techniques while giving consistent posterior distributions, give different posterior weights to different parts of the sky.
    }
    \label{fig:151012_extrinsic}
\end{figure}

\begin{table}[]
    \centering
    \begin{tabular}{c|c|c}
         PSD & No marg & Marg \\ \hline\hline
         Mean & 9.98 & 9.89 \\
         Median & 10.39 & 10.16
    \end{tabular}
    \caption{
    Values of the (natural) log coherent vs incoherent Bayes factor ($\ln$ BCI) for different PSD estimates and likelihoods.
    In both cases the difference between the $\ln$ BCI with and without marginalising over the PSD uncertainty is within typical uncertainties due to finite sampling.
    }
    \label{tab:bci-151012}
\end{table}

\begin{table}[]
    \centering
    \begin{tabular}{c|c}
         PSD & Marg vs no marg \\ \hline\hline
         Mean & 19.26 \\
         Median & 91.67
    \end{tabular}
    \caption{
    Values of the (natural) log Bayes factor comparing the marginalised and unmarginalised likelihood hypotheses.
    In both cases there is a strong preference for the marginalised likelihood better describing the data than the unmarginalised likelihood.
    }
    \label{tab:noise-evidence-151012}
\end{table}

We analyze the data using Bayesian inference as described in Section~\ref{sec:inference} twice, once each with the mean-averaged and median-averaged PSDs to obtain samples from the posterior distribution and Bayesian evidences under four sets of assumptions.
\begin{enumerate}
    \item The data are well described by the mean-estimated PSD and the Whittle likelihood.
    \item The data are well described by the mean-estimated PSD and the Student-Rayleigh likelihood.
    \item The data are well described by the median-estimated PSD and the Whittle likelihood.
    \item The data are well described by the median-estimated PSD and the median marginalized likelihood, Equation~\ref{eq:analytic-median-marginalised}.
\end{enumerate}

In Table~\ref{tab:bci-151012} we show the natural logarithm of the BCI under these four set of assumptions.
We find that both PSD estimation methods have $\ln {\rm BCI} \approx 10$ which is a moderately strong preference for the coherent hypothesis, although we note that a full treatment requires careful consideration of prior odds.
For both PSD estimates the BCI decreases slightly when marginalizing over the uncertainty.
The increase in the BCI when using the median estimated PSD is mirrored in the increased signal-to-noise ratio $\rho$ in the lower panel of Figure~\ref{fig:151012_intrinsic}.
This is likely due to the different handling of non-Gaussian features in the mean and median PSD estimation methods.

In Table~\ref{tab:noise-evidence-151012} we show the natural log Bayes factors comparing the marginalized to unmarginalized likelihoods for both PSD estimation methods.
In both cases, we see a strong preference for the model which marginalizes over the uncertainty.
This preference is much larger for the median PSD estimate.
This can be understood by the fact that the large $|\nu|$ tail of the median marginalized likelihood is broader than the mean marginalized likelihood, c.f., Figure~\ref{fig:data-whitening}, lower-right panel.

In Figures~\ref{fig:151012_intrinsic} and~\ref{fig:151012_extrinsic} we show selected posterior distribution under our four sets of assumptions.
In the top panel of Figure~\ref{fig:151012_intrinsic} we show the posterior distribution for the best measured combination of the component masses, the chirp mass
\begin{equation}
    \mathcal{M} = \frac{(m_1 m_2)^{3/5}}{(m_1 + m_1)^{1/5}}.
\end{equation}
The parameters $m_i$ are the masses of the two component black holes.
In the bottom panel we show the network matched filter signal-to-noise ratio (SNR).
We find that the recovered SNR is larger when using the median PSD estimate and the marginalizing over the uncertainty in the PSD increases the posterior support at SNR less than the maximum found SNR but does not decrease the maximum SNR.
Correspondingly, we see that the posterior for chirp mass is slightly less strongly peaked when marginalizing over uncertainty in the PSD, and when using the median PSD estimate.

In Figure~\ref{fig:151012_extrinsic} we show the posterior distribution for the parameters describing the position on the sky, right ascension $\alpha$ and declination $\delta$.
The impact of marginalizing over the uncertainty in the PSD does not significantly affect the inferred sky localisation of the binary.
However, the two PSD estimation methods recover different posterior distributions within the same region on the sky.

The fact that the differences in the posterior distributions and Bayes factors when using the different PSD estimation methods are larger than the corrections due to marginalizing over the statistical uncertainty mean that either one or both of the estimation methods are producing a biased estimate of the true PSDs.
The source of this bias is presumably non-stationarity and/or non-Gaussianity in the data used to estimate the PSD.
It is not possible to determine which estimate is less biased and so this must be considered as an extra source of systematic uncertainty in off-source PSD estimates.

\section{Discussion}\label{sec:discussion}

Performing astrophysical inference on gravitational-wave data requires an estimate of the noise power spectral density (PSD).
In practice it is not possible to know this perfectly, only to estimate it from the data.
Multiple methods of estimating the PSD are used, and each carries with it a different class of statistical uncertainty.
In this work, we derived the relevant statistical uncertainty for an estimation method, which is widely used when analyzing gravitational-wave transients, the median average.
We obtained a closed-form expression for the likelihood, which marginalizes over the statistical uncertainty, and demonstrated that simulated Gaussian data matches this distribution.

We then applied our new results to the lowest significance transient in the first LIGO/Virgo gravitational-wave transient catalog, GW151012.
We analyzed this event using two different PSDs with likelihoods, which did and did not marginalize over the appropriate statistical uncertainty, one using a median average, and one using a mean average.
We showed the PSD estimation method has a clear effect on the inferred posterior probability distribution and Bayesian evidence.
The changes in the posterior distributions and Bayesian evidence when marginalizing over the statistical uncertainty is more subtle.
However, for applications which require precise estimates of the evidence such as~\cite{Isi2018, Smith2018, Ashton2019b}, these small differences will be crucial.

There are many interesting extensions to the work presented here, which are left to future work.
These include implementing data quality tests when analyzing real gravitational-wave data, which are known to be non-Gaussian and non-stationary over timescales of minutes to hours.
Examples of this can be found in other areas of gravitational-wave data analysis.
For example, searches for gravitational-wave transient signals implement methods to track and mitigate non-stationarity~\cite{Venumadhav2019} and remove large non-Gaussian transients~\cite{Venumadhav2019, Sachdev2019}.
Searches for continuous gravitational-wave sources and the stochastic gravitational-wave background include algorithms to detect and remove stretches of data where the PSD is rapidly fluctuating~\cite{O1Stochastic} or frequencies where the data are known to be non-Gaussian, e.g., around the frequency of AC mains electricity~\cite{LIGOData}.
By combining these methods and the statistical models presented here, we can enable precision astrophysical inference for gravitational-wave transients without large computational overheads.

\section*{Acknowledgements}

We thank Sharan Banagiri, Sylvia Biscoveanu, Katerina Chatziioannou, Pat Meyers, and Joe Romano for helpful discussions and comments on the manuscript.
CT and ET are supported by the Australian Research Council (ARC) CE170100004.
CT acknowledges the support of the National Science Foundation, and the LIGO Laboratory.
ET is supported by ARC FT150100281.
This research has made use of data, software and/or web tools obtained from the Gravitational Wave Open Science Center~\cite{Vallisneri2015, OpenData} (\href{https://www.gw-openscience.org}{https://www.gw-openscience.org}), a service of LIGO Laboratory, the LIGO Scientific Collaboration and the Virgo Collaboration. LIGO is funded by the U.S. National Science Foundation. Virgo is funded by the French Centre National de Recherche Scientifique (CNRS), the Italian Istituto Nazionale della Fisica Nucleare (INFN) and the Dutch Nikhef, with contributions by Polish and Hungarian institutes.
The authors are grateful for computational resources provided by the LIGO Lab and supported by National Science Foundation Grants PHY-0757058 and PHY-0823459.

\bibliography{marg}

\begin{thebibliography}{39}%
\makeatletter
\providecommand \@ifxundefined [1]{%
 \@ifx{#1\undefined}
}%
\providecommand \@ifnum [1]{%
 \ifnum #1\expandafter \@firstoftwo
 \else \expandafter \@secondoftwo
 \fi
}%
\providecommand \@ifx [1]{%
 \ifx #1\expandafter \@firstoftwo
 \else \expandafter \@secondoftwo
 \fi
}%
\providecommand \natexlab [1]{#1}%
\providecommand \enquote  [1]{``#1''}%
\providecommand \bibnamefont  [1]{#1}%
\providecommand \bibfnamefont [1]{#1}%
\providecommand \citenamefont [1]{#1}%
\providecommand \href@noop [0]{\@secondoftwo}%
\providecommand \href [0]{\begingroup \@sanitize@url \@href}%
\providecommand \@href[1]{\@@startlink{#1}\@@href}%
\providecommand \@@href[1]{\endgroup#1\@@endlink}%
\providecommand \@sanitize@url [0]{\catcode `\\12\catcode `\$12\catcode
  `\&12\catcode `\#12\catcode `\^12\catcode `\_12\catcode `\%12\relax}%
\providecommand \@@startlink[1]{}%
\providecommand \@@endlink[0]{}%
\providecommand \url  [0]{\begingroup\@sanitize@url \@url }%
\providecommand \@url [1]{\endgroup\@href {#1}{\urlprefix }}%
\providecommand \urlprefix  [0]{URL }%
\providecommand \Eprint [0]{\href }%
\providecommand \doibase [0]{http://dx.doi.org/}%
\providecommand \selectlanguage [0]{\@gobble}%
\providecommand \bibinfo  [0]{\@secondoftwo}%
\providecommand \bibfield  [0]{\@secondoftwo}%
\providecommand \translation [1]{[#1]}%
\providecommand \BibitemOpen [0]{}%
\providecommand \bibitemStop [0]{}%
\providecommand \bibitemNoStop [0]{.\EOS\space}%
\providecommand \EOS [0]{\spacefactor3000\relax}%
\providecommand \BibitemShut  [1]{\csname bibitem#1\endcsname}%
\let\auto@bib@innerbib\@empty
\bibitem [{\citenamefont {Abbott}\ \emph {et~al.}()\citenamefont {Abbott} \emph
  {et~al.}}]{LIGOData}%
  \BibitemOpen
  \bibfield  {author} {\bibinfo {author} {\bibfnamefont {B.~P.}\ \bibnamefont
  {Abbott}} \emph {et~al.},\ }\href {https://arxiv.org/abs/1908.11170} {\
  }\Eprint {http://arxiv.org/abs/1908.11170} {arXiv:1908.11170} \BibitemShut
  {NoStop}%
\bibitem [{\citenamefont {Romano}\ and\ \citenamefont
  {Cornish}(2017)}]{Romano2017}%
  \BibitemOpen
  \bibfield  {author} {\bibinfo {author} {\bibfnamefont {J.~D.}\ \bibnamefont
  {Romano}}\ and\ \bibinfo {author} {\bibfnamefont {N.~J.}\ \bibnamefont
  {Cornish}},\ }\href {\doibase 10.1007/s41114-017-0004-1} {\bibfield
  {journal} {\bibinfo  {journal} {Living Rev. Relativ.}\ }\textbf {\bibinfo
  {volume} {20}},\ \bibinfo {pages} {2} (\bibinfo {year} {2017})},\ \Eprint
  {http://arxiv.org/abs/1608.06889} {arXiv:1608.06889} \BibitemShut {NoStop}%
\bibitem [{\citenamefont {Littenberg}\ and\ \citenamefont
  {Cornish}(2015)}]{Littenberg2015}%
  \BibitemOpen
  \bibfield  {author} {\bibinfo {author} {\bibfnamefont {T.~B.}\ \bibnamefont
  {Littenberg}}\ and\ \bibinfo {author} {\bibfnamefont {N.~J.}\ \bibnamefont
  {Cornish}},\ }\href {\doibase 10.1103/PhysRevD.91.084034} {\bibfield
  {journal} {\bibinfo  {journal} {Phys. Rev. D}\ }\textbf {\bibinfo {volume}
  {91}},\ \bibinfo {pages} {084034} (\bibinfo {year} {2015})},\ \Eprint
  {http://arxiv.org/abs/1410.3852} {arXiv:1410.3852} \BibitemShut {NoStop}%
\bibitem [{\citenamefont {Welch}(1967)}]{Welch1967}%
  \BibitemOpen
  \bibfield  {author} {\bibinfo {author} {\bibfnamefont {P.}~\bibnamefont
  {Welch}},\ }\href {\doibase 10.1109/TAU.1967.1161901} {\bibfield  {journal}
  {\bibinfo  {journal} {IEEE Trans. Audio Electroacoust.}\ }\textbf {\bibinfo
  {volume} {15}},\ \bibinfo {pages} {70} (\bibinfo {year} {1967})}\BibitemShut
  {NoStop}%
\bibitem [{\citenamefont {{Abbott}}\ \emph {et~al.}(2017)\citenamefont
  {{Abbott}} \emph {et~al.}}]{O1Stochastic}%
  \BibitemOpen
  \bibfield  {author} {\bibinfo {author} {\bibfnamefont {B.~P.}\ \bibnamefont
  {{Abbott}}} \emph {et~al.},\ }\href {\doibase 10.1103/PhysRevLett.118.121101}
  {\bibfield  {journal} {\bibinfo  {journal} {\prl}\ }\textbf {\bibinfo
  {volume} {118}},\ \bibinfo {eid} {121101} (\bibinfo {year} {2017})},\ \Eprint
  {http://arxiv.org/abs/1612.02029} {arXiv:1612.02029 [gr-qc]} \BibitemShut
  {NoStop}%
\bibitem [{\citenamefont {{Usman}}\ \emph {et~al.}(2016)\citenamefont
  {{Usman}}, \citenamefont {{Nitz}}, \citenamefont {{Harry}}, \citenamefont
  {{Biwer}}, \citenamefont {{Brown}}, \citenamefont {{Cabero}}, \citenamefont
  {{Capano}}, \citenamefont {{Dal Canton}}, \citenamefont {{Dent}},
  \citenamefont {{Fairhurst}}, \citenamefont {{Kehl}}, \citenamefont
  {{Keppel}}, \citenamefont {{Krishnan}}, \citenamefont {{Lenon}},
  \citenamefont {{Lundgren}}, \citenamefont {{Nielsen}}, \citenamefont
  {{Pekowsky}}, \citenamefont {{Pfeiffer}}, \citenamefont {{Saulson}},
  \citenamefont {{West}},\ and\ \citenamefont {{Willis}}}]{Usman2016}%
  \BibitemOpen
  \bibfield  {author} {\bibinfo {author} {\bibfnamefont {S.~A.}\ \bibnamefont
  {{Usman}}}, \bibinfo {author} {\bibfnamefont {A.~H.}\ \bibnamefont {{Nitz}}},
  \bibinfo {author} {\bibfnamefont {I.~W.}\ \bibnamefont {{Harry}}}, \bibinfo
  {author} {\bibfnamefont {C.~M.}\ \bibnamefont {{Biwer}}}, \bibinfo {author}
  {\bibfnamefont {D.~A.}\ \bibnamefont {{Brown}}}, \bibinfo {author}
  {\bibfnamefont {M.}~\bibnamefont {{Cabero}}}, \bibinfo {author}
  {\bibfnamefont {C.~D.}\ \bibnamefont {{Capano}}}, \bibinfo {author}
  {\bibfnamefont {T.}~\bibnamefont {{Dal Canton}}}, \bibinfo {author}
  {\bibfnamefont {T.}~\bibnamefont {{Dent}}}, \bibinfo {author} {\bibfnamefont
  {S.}~\bibnamefont {{Fairhurst}}}, \bibinfo {author} {\bibfnamefont {M.~S.}\
  \bibnamefont {{Kehl}}}, \bibinfo {author} {\bibfnamefont {D.}~\bibnamefont
  {{Keppel}}}, \bibinfo {author} {\bibfnamefont {B.}~\bibnamefont
  {{Krishnan}}}, \bibinfo {author} {\bibfnamefont {A.}~\bibnamefont {{Lenon}}},
  \bibinfo {author} {\bibfnamefont {A.}~\bibnamefont {{Lundgren}}}, \bibinfo
  {author} {\bibfnamefont {A.~B.}\ \bibnamefont {{Nielsen}}}, \bibinfo {author}
  {\bibfnamefont {L.~P.}\ \bibnamefont {{Pekowsky}}}, \bibinfo {author}
  {\bibfnamefont {H.~P.}\ \bibnamefont {{Pfeiffer}}}, \bibinfo {author}
  {\bibfnamefont {P.~R.}\ \bibnamefont {{Saulson}}}, \bibinfo {author}
  {\bibfnamefont {M.}~\bibnamefont {{West}}}, \ and\ \bibinfo {author}
  {\bibfnamefont {J.~L.}\ \bibnamefont {{Willis}}},\ }\href {\doibase
  10.1088/0264-9381/33/21/215004} {\bibfield  {journal} {\bibinfo  {journal}
  {Classical and Quantum Gravity}\ }\textbf {\bibinfo {volume} {33}},\ \bibinfo
  {eid} {215004} (\bibinfo {year} {2016})},\ \Eprint
  {http://arxiv.org/abs/1508.02357} {arXiv:1508.02357 [gr-qc]} \BibitemShut
  {NoStop}%
\bibitem [{\citenamefont {{Pankow}}\ \emph {et~al.}(2018)\citenamefont
  {{Pankow}}, \citenamefont {{Chatziioannou}}, \citenamefont {{Chase}},
  \citenamefont {{Littenberg}}, \citenamefont {{Evans}}, \citenamefont
  {{McIver}}, \citenamefont {{Cornish}}, \citenamefont {{Haster}},
  \citenamefont {{Kanner}}, \citenamefont {{Raymond}}, \citenamefont
  {{Vitale}},\ and\ \citenamefont {{Zimmerman}}}]{Pankow2018}%
  \BibitemOpen
  \bibfield  {author} {\bibinfo {author} {\bibfnamefont {C.}~\bibnamefont
  {{Pankow}}}, \bibinfo {author} {\bibfnamefont {K.}~\bibnamefont
  {{Chatziioannou}}}, \bibinfo {author} {\bibfnamefont {E.~A.}\ \bibnamefont
  {{Chase}}}, \bibinfo {author} {\bibfnamefont {T.~B.}\ \bibnamefont
  {{Littenberg}}}, \bibinfo {author} {\bibfnamefont {M.}~\bibnamefont
  {{Evans}}}, \bibinfo {author} {\bibfnamefont {J.}~\bibnamefont {{McIver}}},
  \bibinfo {author} {\bibfnamefont {N.~J.}\ \bibnamefont {{Cornish}}}, \bibinfo
  {author} {\bibfnamefont {C.-J.}\ \bibnamefont {{Haster}}}, \bibinfo {author}
  {\bibfnamefont {J.}~\bibnamefont {{Kanner}}}, \bibinfo {author}
  {\bibfnamefont {V.}~\bibnamefont {{Raymond}}}, \bibinfo {author}
  {\bibfnamefont {S.}~\bibnamefont {{Vitale}}}, \ and\ \bibinfo {author}
  {\bibfnamefont {A.}~\bibnamefont {{Zimmerman}}},\ }\href {\doibase
  10.1103/PhysRevD.98.084016} {\bibfield  {journal} {\bibinfo  {journal}
  {\prd}\ }\textbf {\bibinfo {volume} {98}},\ \bibinfo {eid} {084016} (\bibinfo
  {year} {2018})},\ \Eprint {http://arxiv.org/abs/1808.03619} {arXiv:1808.03619
  [gr-qc]} \BibitemShut {NoStop}%
\bibitem [{\citenamefont {{Driggers}}\ \emph {et~al.}(2019)\citenamefont
  {{Driggers}}, \citenamefont {{Vitale}}, \citenamefont {{Lundgren}},
  \citenamefont {{Evans}}, \citenamefont {{Kawabe}}, \citenamefont {{Dwyer}},
  \citenamefont {{Izumi}}, \citenamefont {{Schofield}}, \citenamefont
  {{Effler}}, \citenamefont {{Sigg}} \emph {et~al.}}]{Driggers2019}%
  \BibitemOpen
  \bibfield  {author} {\bibinfo {author} {\bibfnamefont {J.~C.}\ \bibnamefont
  {{Driggers}}}, \bibinfo {author} {\bibfnamefont {S.}~\bibnamefont
  {{Vitale}}}, \bibinfo {author} {\bibfnamefont {A.~P.}\ \bibnamefont
  {{Lundgren}}}, \bibinfo {author} {\bibfnamefont {M.}~\bibnamefont {{Evans}}},
  \bibinfo {author} {\bibfnamefont {K.}~\bibnamefont {{Kawabe}}}, \bibinfo
  {author} {\bibfnamefont {S.~E.}\ \bibnamefont {{Dwyer}}}, \bibinfo {author}
  {\bibfnamefont {K.}~\bibnamefont {{Izumi}}}, \bibinfo {author} {\bibfnamefont
  {R.~M.~S.}\ \bibnamefont {{Schofield}}}, \bibinfo {author} {\bibfnamefont
  {A.}~\bibnamefont {{Effler}}}, \bibinfo {author} {\bibfnamefont
  {D.}~\bibnamefont {{Sigg}}},  \emph {et~al.},\ }\href {\doibase
  10.1103/PhysRevD.99.042001} {\bibfield  {journal} {\bibinfo  {journal}
  {\prd}\ }\textbf {\bibinfo {volume} {99}},\ \bibinfo {eid} {042001} (\bibinfo
  {year} {2019})},\ \Eprint {http://arxiv.org/abs/1806.00532} {arXiv:1806.00532
  [astro-ph.IM]} \BibitemShut {NoStop}%
\bibitem [{\citenamefont {{Davis}}\ \emph {et~al.}(2019)\citenamefont
  {{Davis}}, \citenamefont {{Massinger}}, \citenamefont {{Lundgren}},
  \citenamefont {{Driggers}}, \citenamefont {{Urban}},\ and\ \citenamefont
  {{Nuttall}}}]{Davis2019}%
  \BibitemOpen
  \bibfield  {author} {\bibinfo {author} {\bibfnamefont {D.}~\bibnamefont
  {{Davis}}}, \bibinfo {author} {\bibfnamefont {T.}~\bibnamefont
  {{Massinger}}}, \bibinfo {author} {\bibfnamefont {A.}~\bibnamefont
  {{Lundgren}}}, \bibinfo {author} {\bibfnamefont {J.~C.}\ \bibnamefont
  {{Driggers}}}, \bibinfo {author} {\bibfnamefont {A.~L.}\ \bibnamefont
  {{Urban}}}, \ and\ \bibinfo {author} {\bibfnamefont {L.}~\bibnamefont
  {{Nuttall}}},\ }\href {\doibase 10.1088/1361-6382/ab01c5} {\bibfield
  {journal} {\bibinfo  {journal} {Classical and Quantum Gravity}\ }\textbf
  {\bibinfo {volume} {36}},\ \bibinfo {eid} {055011} (\bibinfo {year}
  {2019})},\ \Eprint {http://arxiv.org/abs/1809.05348} {arXiv:1809.05348
  [astro-ph.IM]} \BibitemShut {NoStop}%
\bibitem [{\citenamefont {{Sachdev}}\ \emph {et~al.}(2019)\citenamefont
  {{Sachdev}}, \citenamefont {{Caudill}}, \citenamefont {{Fong}}, \citenamefont
  {{Lo}}, \citenamefont {{Messick}}, \citenamefont {{Mukherjee}}, \citenamefont
  {{Magee}}, \citenamefont {{Tsukada}}, \citenamefont {{Blackburn}},
  \citenamefont {{Brady}}, \citenamefont {{Brockill}} \emph
  {et~al.}}]{Sachdev2019}%
  \BibitemOpen
  \bibfield  {author} {\bibinfo {author} {\bibfnamefont {S.}~\bibnamefont
  {{Sachdev}}}, \bibinfo {author} {\bibfnamefont {S.}~\bibnamefont
  {{Caudill}}}, \bibinfo {author} {\bibfnamefont {H.}~\bibnamefont {{Fong}}},
  \bibinfo {author} {\bibfnamefont {R.~K.~L.}\ \bibnamefont {{Lo}}}, \bibinfo
  {author} {\bibfnamefont {C.}~\bibnamefont {{Messick}}}, \bibinfo {author}
  {\bibfnamefont {D.}~\bibnamefont {{Mukherjee}}}, \bibinfo {author}
  {\bibfnamefont {R.}~\bibnamefont {{Magee}}}, \bibinfo {author} {\bibfnamefont
  {L.}~\bibnamefont {{Tsukada}}}, \bibinfo {author} {\bibfnamefont
  {K.}~\bibnamefont {{Blackburn}}}, \bibinfo {author} {\bibfnamefont
  {P.}~\bibnamefont {{Brady}}}, \bibinfo {author} {\bibfnamefont
  {P.}~\bibnamefont {{Brockill}}},  \emph {et~al.},\ }\href@noop {} {\
  (\bibinfo {year} {2019})},\ \Eprint {http://arxiv.org/abs/1901.08580}
  {arXiv:1901.08580 [gr-qc]} \BibitemShut {NoStop}%
\bibitem [{\citenamefont {{Venumadhav}}\ \emph {et~al.}(2019)\citenamefont
  {{Venumadhav}}, \citenamefont {{Zackay}}, \citenamefont {{Roulet}},
  \citenamefont {{Dai}},\ and\ \citenamefont {{Zaldarriaga}}}]{Venumadhav2019}%
  \BibitemOpen
  \bibfield  {author} {\bibinfo {author} {\bibfnamefont {T.}~\bibnamefont
  {{Venumadhav}}}, \bibinfo {author} {\bibfnamefont {B.}~\bibnamefont
  {{Zackay}}}, \bibinfo {author} {\bibfnamefont {J.}~\bibnamefont {{Roulet}}},
  \bibinfo {author} {\bibfnamefont {L.}~\bibnamefont {{Dai}}}, \ and\ \bibinfo
  {author} {\bibfnamefont {M.}~\bibnamefont {{Zaldarriaga}}},\ }\href {\doibase
  10.1103/PhysRevD.100.023011} {\bibfield  {journal} {\bibinfo  {journal}
  {\prd}\ }\textbf {\bibinfo {volume} {100}},\ \bibinfo {eid} {023011}
  (\bibinfo {year} {2019})},\ \Eprint {http://arxiv.org/abs/1902.10341}
  {arXiv:1902.10341 [astro-ph.IM]} \BibitemShut {NoStop}%
\bibitem [{\citenamefont {{Vajente}}\ \emph {et~al.}(2020)\citenamefont
  {{Vajente}}, \citenamefont {{Huang}}, \citenamefont {{Isi}}, \citenamefont
  {{Driggers}}, \citenamefont {{Kissel}}, \citenamefont {{Szczepa{\'n}czyk}},\
  and\ \citenamefont {{Vitale}}}]{Vajente2020}%
  \BibitemOpen
  \bibfield  {author} {\bibinfo {author} {\bibfnamefont {G.}~\bibnamefont
  {{Vajente}}}, \bibinfo {author} {\bibfnamefont {Y.}~\bibnamefont {{Huang}}},
  \bibinfo {author} {\bibfnamefont {M.}~\bibnamefont {{Isi}}}, \bibinfo
  {author} {\bibfnamefont {J.~C.}\ \bibnamefont {{Driggers}}}, \bibinfo
  {author} {\bibfnamefont {J.~S.}\ \bibnamefont {{Kissel}}}, \bibinfo {author}
  {\bibfnamefont {M.~J.}\ \bibnamefont {{Szczepa{\'n}czyk}}}, \ and\ \bibinfo
  {author} {\bibfnamefont {S.}~\bibnamefont {{Vitale}}},\ }\href {\doibase
  10.1103/PhysRevD.101.042003} {\bibfield  {journal} {\bibinfo  {journal}
  {\prd}\ }\textbf {\bibinfo {volume} {101}},\ \bibinfo {eid} {042003}
  (\bibinfo {year} {2020})},\ \Eprint {http://arxiv.org/abs/1911.09083}
  {arXiv:1911.09083 [gr-qc]} \BibitemShut {NoStop}%
\bibitem [{\citenamefont {Chatziioannou}\ \emph {et~al.}(2019)\citenamefont
  {Chatziioannou}, \citenamefont {Haster}, \citenamefont {Littenberg},
  \citenamefont {Farr}, \citenamefont {Ghonge}, \citenamefont {Millhouse},
  \citenamefont {Clark},\ and\ \citenamefont {Cornish}}]{Chatziioannou2019}%
  \BibitemOpen
  \bibfield  {author} {\bibinfo {author} {\bibfnamefont {K.}~\bibnamefont
  {Chatziioannou}}, \bibinfo {author} {\bibfnamefont {C.-J.}\ \bibnamefont
  {Haster}}, \bibinfo {author} {\bibfnamefont {T.~B.}\ \bibnamefont
  {Littenberg}}, \bibinfo {author} {\bibfnamefont {W.~M.}\ \bibnamefont
  {Farr}}, \bibinfo {author} {\bibfnamefont {S.}~\bibnamefont {Ghonge}},
  \bibinfo {author} {\bibfnamefont {M.}~\bibnamefont {Millhouse}}, \bibinfo
  {author} {\bibfnamefont {J.~A.}\ \bibnamefont {Clark}}, \ and\ \bibinfo
  {author} {\bibfnamefont {N.}~\bibnamefont {Cornish}},\ }\href
  {http://arxiv.org/abs/1907.06540} {\  (\bibinfo {year} {2019})},\ \Eprint
  {http://arxiv.org/abs/1907.06540} {arXiv:1907.06540} \BibitemShut {NoStop}%
\bibitem [{\citenamefont {R{\"{o}}ver}(2011)}]{Rover2011a}%
  \BibitemOpen
  \bibfield  {author} {\bibinfo {author} {\bibfnamefont {C.}~\bibnamefont
  {R{\"{o}}ver}},\ }\href {\doibase 10.1103/PhysRevD.84.122004} {\bibfield
  {journal} {\bibinfo  {journal} {Phys. Rev. D}\ }\textbf {\bibinfo {volume}
  {84}},\ \bibinfo {pages} {122004} (\bibinfo {year} {2011})},\ \Eprint
  {http://arxiv.org/abs/1109.0442} {arXiv:1109.0442} \BibitemShut {NoStop}%
\bibitem [{\citenamefont {{R{\"o}ver}}\ \emph {et~al.}(2011)\citenamefont
  {{R{\"o}ver}}, \citenamefont {{Meyer}},\ and\ \citenamefont
  {{Christensen}}}]{Rover2011b}%
  \BibitemOpen
  \bibfield  {author} {\bibinfo {author} {\bibfnamefont {C.}~\bibnamefont
  {{R{\"o}ver}}}, \bibinfo {author} {\bibfnamefont {R.}~\bibnamefont
  {{Meyer}}}, \ and\ \bibinfo {author} {\bibfnamefont {N.}~\bibnamefont
  {{Christensen}}},\ }\href {\doibase 10.1088/0264-9381/28/1/015010} {\bibfield
   {journal} {\bibinfo  {journal} {Classical and Quantum Gravity}\ }\textbf
  {\bibinfo {volume} {28}},\ \bibinfo {eid} {015010} (\bibinfo {year}
  {2011})},\ \Eprint {http://arxiv.org/abs/0804.3853} {arXiv:0804.3853
  [stat.ME]} \BibitemShut {NoStop}%
\bibitem [{\citenamefont {Yamamoto}\ \emph {et~al.}(2016)\citenamefont
  {Yamamoto}, \citenamefont {Hayama}, \citenamefont {Mano}, \citenamefont
  {Itoh},\ and\ \citenamefont {Kanda}}]{Yamamoto2016}%
  \BibitemOpen
  \bibfield  {author} {\bibinfo {author} {\bibfnamefont {T.}~\bibnamefont
  {Yamamoto}}, \bibinfo {author} {\bibfnamefont {K.}~\bibnamefont {Hayama}},
  \bibinfo {author} {\bibfnamefont {S.}~\bibnamefont {Mano}}, \bibinfo {author}
  {\bibfnamefont {Y.}~\bibnamefont {Itoh}}, \ and\ \bibinfo {author}
  {\bibfnamefont {N.}~\bibnamefont {Kanda}},\ }\href {\doibase
  10.1103/PhysRevD.93.082005} {\bibfield  {journal} {\bibinfo  {journal} {Phys.
  Rev. D}\ }\textbf {\bibinfo {volume} {93}},\ \bibinfo {pages} {082005}
  (\bibinfo {year} {2016})}\BibitemShut {NoStop}%
\bibitem [{\citenamefont {Banagiri}\ \emph {et~al.}(2019)\citenamefont
  {Banagiri}, \citenamefont {Coughlin}, \citenamefont {Clark}, \citenamefont
  {Lasky}, \citenamefont {Bizouard}, \citenamefont {Talbot}, \citenamefont
  {Thrane},\ and\ \citenamefont {Mandic}}]{Banagiri2019}%
  \BibitemOpen
  \bibfield  {author} {\bibinfo {author} {\bibfnamefont {S.}~\bibnamefont
  {Banagiri}}, \bibinfo {author} {\bibfnamefont {M.~W.}\ \bibnamefont
  {Coughlin}}, \bibinfo {author} {\bibfnamefont {J.}~\bibnamefont {Clark}},
  \bibinfo {author} {\bibfnamefont {P.~D.}\ \bibnamefont {Lasky}}, \bibinfo
  {author} {\bibfnamefont {M.~A.}\ \bibnamefont {Bizouard}}, \bibinfo {author}
  {\bibfnamefont {C.}~\bibnamefont {Talbot}}, \bibinfo {author} {\bibfnamefont
  {E.}~\bibnamefont {Thrane}}, \ and\ \bibinfo {author} {\bibfnamefont
  {V.}~\bibnamefont {Mandic}},\ }\href {http://arxiv.org/abs/1909.01934} {\
  (\bibinfo {year} {2019})},\ \Eprint {http://arxiv.org/abs/1909.01934}
  {arXiv:1909.01934} \BibitemShut {NoStop}%
\bibitem [{\citenamefont {Cornish}\ and\ \citenamefont
  {Littenberg}(2015)}]{Cornish2015}%
  \BibitemOpen
  \bibfield  {author} {\bibinfo {author} {\bibfnamefont {N.~J.}\ \bibnamefont
  {Cornish}}\ and\ \bibinfo {author} {\bibfnamefont {T.~B.}\ \bibnamefont
  {Littenberg}},\ }\href {\doibase 10.1088/0264-9381/32/13/135012} {\bibfield
  {journal} {\bibinfo  {journal} {Class. Quantum Gravity}\ }\textbf {\bibinfo
  {volume} {32}},\ \bibinfo {pages} {135012} (\bibinfo {year} {2015})},\
  \Eprint {http://arxiv.org/abs/1410.3835} {arXiv:1410.3835} \BibitemShut
  {NoStop}%
\bibitem [{\citenamefont {{Abbott}}\ \emph {et~al.}(2019)\citenamefont
  {{Abbott}} \emph {et~al.}}]{GWTC1}%
  \BibitemOpen
  \bibfield  {author} {\bibinfo {author} {\bibfnamefont {B.~P.}\ \bibnamefont
  {{Abbott}}} \emph {et~al.},\ }\href {\doibase 10.1103/PhysRevX.9.031040}
  {\bibfield  {journal} {\bibinfo  {journal} {Physical Review X}\ }\textbf
  {\bibinfo {volume} {9}},\ \bibinfo {eid} {031040} (\bibinfo {year} {2019})},\
  \Eprint {http://arxiv.org/abs/1811.12907} {arXiv:1811.12907 [astro-ph.HE]}
  \BibitemShut {NoStop}%
\bibitem [{\citenamefont {{Biscoveanu}}\ \emph {et~al.}(2020)\citenamefont
  {{Biscoveanu}}, \citenamefont {{Haster}}, \citenamefont {{Vitale}},\ and\
  \citenamefont {{Davies}}}]{Biscoveanu2020}%
  \BibitemOpen
  \bibfield  {author} {\bibinfo {author} {\bibfnamefont {S.}~\bibnamefont
  {{Biscoveanu}}}, \bibinfo {author} {\bibfnamefont {C.-J.}\ \bibnamefont
  {{Haster}}}, \bibinfo {author} {\bibfnamefont {S.}~\bibnamefont {{Vitale}}},
  \ and\ \bibinfo {author} {\bibfnamefont {J.}~\bibnamefont {{Davies}}},\
  }\href@noop {} {\bibfield  {journal} {\bibinfo  {journal} {arXiv e-prints}\ }
  (\bibinfo {year} {2020})},\ \Eprint {http://arxiv.org/abs/2004.05149}
  {arXiv:2004.05149 [astro-ph.HE]} \BibitemShut {NoStop}%
\bibitem [{\citenamefont {Talbot}(2020)}]{Talbot2020}%
  \BibitemOpen
  \bibfield  {author} {\bibinfo {author} {\bibfnamefont {C.}~\bibnamefont
  {Talbot}},\ }\emph {\bibinfo {title} {Astrophysics of Binary Black Holes at
  the Dawn of Gravitational-Wave Astronomy}},\ \href {\doibase
  10.26180/5e61a9fc39b73} {Ph.D. thesis},\ \bibinfo  {school} {Monash
  University} (\bibinfo {year} {2020})\BibitemShut {NoStop}%
\bibitem [{\citenamefont {Allen}\ \emph {et~al.}(2012)\citenamefont {Allen},
  \citenamefont {Anderson}, \citenamefont {Brady}, \citenamefont {Brown},\ and\
  \citenamefont {Creighton}}]{Allen2012}%
  \BibitemOpen
  \bibfield  {author} {\bibinfo {author} {\bibfnamefont {B.}~\bibnamefont
  {Allen}}, \bibinfo {author} {\bibfnamefont {W.~G.}\ \bibnamefont {Anderson}},
  \bibinfo {author} {\bibfnamefont {P.~R.}\ \bibnamefont {Brady}}, \bibinfo
  {author} {\bibfnamefont {D.~A.}\ \bibnamefont {Brown}}, \ and\ \bibinfo
  {author} {\bibfnamefont {J.~D.~E.}\ \bibnamefont {Creighton}},\ }\href
  {\doibase 10.1103/PhysRevD.85.122006} {\bibfield  {journal} {\bibinfo
  {journal} {Phys. Rev. D}\ }\textbf {\bibinfo {volume} {85}},\ \bibinfo
  {pages} {122006} (\bibinfo {year} {2012})},\ \Eprint
  {http://arxiv.org/abs/0509116} {arXiv:0509116 [gr-qc]} \BibitemShut {NoStop}%
\bibitem [{\citenamefont {Veitch}\ \emph {et~al.}(2015)\citenamefont {Veitch},
  \citenamefont {Raymond}, \citenamefont {Farr}, \citenamefont {Farr},
  \citenamefont {Graff}, \citenamefont {Vitale}, \citenamefont {Aylott},
  \citenamefont {Blackburn}, \citenamefont {Christensen}, \citenamefont
  {Coughlin}, \citenamefont {{Del Pozzo}}, \citenamefont {Feroz}, \citenamefont
  {Gair}, \citenamefont {Haster}, \citenamefont {Kalogera}, \citenamefont
  {Littenberg}, \citenamefont {Mandel}, \citenamefont {O'Shaughnessy},
  \citenamefont {Pitkin}, \citenamefont {Rodriguez}, \citenamefont
  {R{\"{o}}ver}, \citenamefont {Sidery}, \citenamefont {Smith}, \citenamefont
  {{Van Der Sluys}}, \citenamefont {Vecchio}, \citenamefont {Vousden},\ and\
  \citenamefont {Wade}}]{Veitch2015}%
  \BibitemOpen
  \bibfield  {author} {\bibinfo {author} {\bibfnamefont {J.}~\bibnamefont
  {Veitch}}, \bibinfo {author} {\bibfnamefont {V.}~\bibnamefont {Raymond}},
  \bibinfo {author} {\bibfnamefont {B.}~\bibnamefont {Farr}}, \bibinfo {author}
  {\bibfnamefont {W.}~\bibnamefont {Farr}}, \bibinfo {author} {\bibfnamefont
  {P.}~\bibnamefont {Graff}}, \bibinfo {author} {\bibfnamefont
  {S.}~\bibnamefont {Vitale}}, \bibinfo {author} {\bibfnamefont
  {B.}~\bibnamefont {Aylott}}, \bibinfo {author} {\bibfnamefont
  {K.}~\bibnamefont {Blackburn}}, \bibinfo {author} {\bibfnamefont
  {N.}~\bibnamefont {Christensen}}, \bibinfo {author} {\bibfnamefont
  {M.}~\bibnamefont {Coughlin}}, \bibinfo {author} {\bibfnamefont
  {W.}~\bibnamefont {{Del Pozzo}}}, \bibinfo {author} {\bibfnamefont
  {F.}~\bibnamefont {Feroz}}, \bibinfo {author} {\bibfnamefont
  {J.}~\bibnamefont {Gair}}, \bibinfo {author} {\bibfnamefont {C.-J.}\
  \bibnamefont {Haster}}, \bibinfo {author} {\bibfnamefont {V.}~\bibnamefont
  {Kalogera}}, \bibinfo {author} {\bibfnamefont {T.}~\bibnamefont
  {Littenberg}}, \bibinfo {author} {\bibfnamefont {I.}~\bibnamefont {Mandel}},
  \bibinfo {author} {\bibfnamefont {R.}~\bibnamefont {O'Shaughnessy}}, \bibinfo
  {author} {\bibfnamefont {M.}~\bibnamefont {Pitkin}}, \bibinfo {author}
  {\bibfnamefont {C.}~\bibnamefont {Rodriguez}}, \bibinfo {author}
  {\bibfnamefont {C.}~\bibnamefont {R{\"{o}}ver}}, \bibinfo {author}
  {\bibfnamefont {T.}~\bibnamefont {Sidery}}, \bibinfo {author} {\bibfnamefont
  {R.}~\bibnamefont {Smith}}, \bibinfo {author} {\bibfnamefont
  {M.}~\bibnamefont {{Van Der Sluys}}}, \bibinfo {author} {\bibfnamefont
  {A.}~\bibnamefont {Vecchio}}, \bibinfo {author} {\bibfnamefont
  {W.}~\bibnamefont {Vousden}}, \ and\ \bibinfo {author} {\bibfnamefont
  {L.}~\bibnamefont {Wade}},\ }\href {\doibase 10.1103/PhysRevD.91.042003}
  {\bibfield  {journal} {\bibinfo  {journal} {Phys. Rev. D}\ }\textbf {\bibinfo
  {volume} {91}},\ \bibinfo {pages} {42003} (\bibinfo {year}
  {2015})}\BibitemShut {NoStop}%
\bibitem [{\citenamefont {{Isi}}\ \emph {et~al.}(2018)\citenamefont {{Isi}},
  \citenamefont {{Smith}}, \citenamefont {{Vitale}}, \citenamefont
  {{Massinger}}, \citenamefont {{Kanner}},\ and\ \citenamefont
  {{Vajpeyi}}}]{Isi2018}%
  \BibitemOpen
  \bibfield  {author} {\bibinfo {author} {\bibfnamefont {M.}~\bibnamefont
  {{Isi}}}, \bibinfo {author} {\bibfnamefont {R.}~\bibnamefont {{Smith}}},
  \bibinfo {author} {\bibfnamefont {S.}~\bibnamefont {{Vitale}}}, \bibinfo
  {author} {\bibfnamefont {T.~J.}\ \bibnamefont {{Massinger}}}, \bibinfo
  {author} {\bibfnamefont {J.}~\bibnamefont {{Kanner}}}, \ and\ \bibinfo
  {author} {\bibfnamefont {A.}~\bibnamefont {{Vajpeyi}}},\ }\href {\doibase
  10.1103/PhysRevD.98.042007} {\bibfield  {journal} {\bibinfo  {journal}
  {\prd}\ }\textbf {\bibinfo {volume} {98}},\ \bibinfo {eid} {042007} (\bibinfo
  {year} {2018})},\ \Eprint {http://arxiv.org/abs/1803.09783} {arXiv:1803.09783
  [gr-qc]} \BibitemShut {NoStop}%
\bibitem [{\citenamefont {{Ashton}}\ \emph {et~al.}(2019)\citenamefont
  {{Ashton}}, \citenamefont {{Thrane}},\ and\ \citenamefont
  {{Smith}}}]{Ashton2019b}%
  \BibitemOpen
  \bibfield  {author} {\bibinfo {author} {\bibfnamefont {G.}~\bibnamefont
  {{Ashton}}}, \bibinfo {author} {\bibfnamefont {E.}~\bibnamefont {{Thrane}}},
  \ and\ \bibinfo {author} {\bibfnamefont {R.~J.~E.}\ \bibnamefont {{Smith}}},\
  }\href {\doibase 10.1103/PhysRevD.100.123018} {\bibfield  {journal} {\bibinfo
   {journal} {\prd}\ }\textbf {\bibinfo {volume} {100}},\ \bibinfo {eid}
  {123018} (\bibinfo {year} {2019})},\ \Eprint
  {http://arxiv.org/abs/1909.11872} {arXiv:1909.11872 [gr-qc]} \BibitemShut
  {NoStop}%
\bibitem [{\citenamefont {{Smith}}\ and\ \citenamefont
  {{Thrane}}(2018)}]{Smith2018}%
  \BibitemOpen
  \bibfield  {author} {\bibinfo {author} {\bibfnamefont {R.}~\bibnamefont
  {{Smith}}}\ and\ \bibinfo {author} {\bibfnamefont {E.}~\bibnamefont
  {{Thrane}}},\ }\href {\doibase 10.1103/PhysRevX.8.021019} {\bibfield
  {journal} {\bibinfo  {journal} {Physical Review X}\ }\textbf {\bibinfo
  {volume} {8}},\ \bibinfo {eid} {021019} (\bibinfo {year} {2018})},\ \Eprint
  {http://arxiv.org/abs/1712.00688} {arXiv:1712.00688 [gr-qc]} \BibitemShut
  {NoStop}%
\bibitem [{\citenamefont {Hastings}(1970)}]{Hastings1970}%
  \BibitemOpen
  \bibfield  {author} {\bibinfo {author} {\bibfnamefont {W.~K.}\ \bibnamefont
  {Hastings}},\ }\href {\doibase 10.1093/biomet/57.1.97} {\bibfield  {journal}
  {\bibinfo  {journal} {Biometrika}\ }\textbf {\bibinfo {volume} {57}},\
  \bibinfo {pages} {97} (\bibinfo {year} {1970})},\ \Eprint
  {http://arxiv.org/abs/https://academic.oup.com/biomet/article-pdf/57/1/97/23940249/57-1-97.pdf}
  {https://academic.oup.com/biomet/article-pdf/57/1/97/23940249/57-1-97.pdf}
  \BibitemShut {NoStop}%
\bibitem [{\citenamefont {Skilling}(2006)}]{Skilling2006}%
  \BibitemOpen
  \bibfield  {author} {\bibinfo {author} {\bibfnamefont {J.}~\bibnamefont
  {Skilling}},\ }\href {\doibase 10.1214/06-BA127} {\bibfield  {journal}
  {\bibinfo  {journal} {Bayesian Anal.}\ }\textbf {\bibinfo {volume} {1}},\
  \bibinfo {pages} {833} (\bibinfo {year} {2006})}\BibitemShut {NoStop}%
\bibitem [{\citenamefont {Thrane}\ and\ \citenamefont
  {Talbot}(2019)}]{Thrane2019}%
  \BibitemOpen
  \bibfield  {author} {\bibinfo {author} {\bibfnamefont {E.}~\bibnamefont
  {Thrane}}\ and\ \bibinfo {author} {\bibfnamefont {C.}~\bibnamefont
  {Talbot}},\ }\href {\doibase 10.1017/pasa.2019.2} {\bibfield  {journal}
  {\bibinfo  {journal} {Publ. Astron. Soc. Aust.}\ }\textbf {\bibinfo {volume}
  {36}},\ \bibinfo {pages} {e010} (\bibinfo {year} {2019})},\ \Eprint
  {http://arxiv.org/abs/1809.02293} {arXiv:1809.02293} \BibitemShut {NoStop}%
\bibitem [{\citenamefont {Payne}\ \emph {et~al.}()\citenamefont {Payne},
  \citenamefont {Talbot},\ and\ \citenamefont {Thrane}}]{Payne2019}%
  \BibitemOpen
  \bibfield  {author} {\bibinfo {author} {\bibfnamefont {E.}~\bibnamefont
  {Payne}}, \bibinfo {author} {\bibfnamefont {C.}~\bibnamefont {Talbot}}, \
  and\ \bibinfo {author} {\bibfnamefont {E.}~\bibnamefont {Thrane}},\ }\href
  {http://arxiv.org/abs/1905.05477} {\ }\Eprint
  {http://arxiv.org/abs/1905.05477} {arXiv:1905.05477} \BibitemShut {NoStop}%
\bibitem [{\citenamefont {Speagle}()}]{Speagle2019}%
  \BibitemOpen
  \bibfield  {author} {\bibinfo {author} {\bibfnamefont {J.~S.}\ \bibnamefont
  {Speagle}},\ }\href {https://https://arxiv.org/abs/1904.02180} {\ }\Eprint
  {http://arxiv.org/abs/1904.02180} {arXiv:1904.02180} \BibitemShut {NoStop}%
\bibitem [{\citenamefont {Ashton}\ \emph {et~al.}(2019)\citenamefont {Ashton},
  \citenamefont {H{\"{u}}bner}, \citenamefont {Lasky}, \citenamefont {Talbot},
  \citenamefont {Ackley}, \citenamefont {Biscoveanu}, \citenamefont {Chu},
  \citenamefont {Divakarla}, \citenamefont {Easter}, \citenamefont {Goncharov}
  \emph {et~al.}}]{Ashton2019a}%
  \BibitemOpen
  \bibfield  {author} {\bibinfo {author} {\bibfnamefont {G.}~\bibnamefont
  {Ashton}}, \bibinfo {author} {\bibfnamefont {M.}~\bibnamefont
  {H{\"{u}}bner}}, \bibinfo {author} {\bibfnamefont {P.~D.}\ \bibnamefont
  {Lasky}}, \bibinfo {author} {\bibfnamefont {C.}~\bibnamefont {Talbot}},
  \bibinfo {author} {\bibfnamefont {K.}~\bibnamefont {Ackley}}, \bibinfo
  {author} {\bibfnamefont {S.}~\bibnamefont {Biscoveanu}}, \bibinfo {author}
  {\bibfnamefont {Q.}~\bibnamefont {Chu}}, \bibinfo {author} {\bibfnamefont
  {A.}~\bibnamefont {Divakarla}}, \bibinfo {author} {\bibfnamefont {P.~J.}\
  \bibnamefont {Easter}}, \bibinfo {author} {\bibfnamefont {B.}~\bibnamefont
  {Goncharov}},  \emph {et~al.},\ }\href {\doibase 10.3847/1538-4365/ab06fc}
  {\bibfield  {journal} {\bibinfo  {journal} {Astrophys. J. Suppl. Ser.}\
  }\textbf {\bibinfo {volume} {241}},\ \bibinfo {pages} {27} (\bibinfo {year}
  {2019})},\ \Eprint {http://arxiv.org/abs/1811.02042} {arXiv:1811.02042}
  \BibitemShut {NoStop}%
\bibitem [{\citenamefont {{H{\"u}bner}}\ \emph {et~al.}(2020)\citenamefont
  {{H{\"u}bner}}, \citenamefont {{Talbot}}, \citenamefont {{Lasky}},\ and\
  \citenamefont {{Thrane}}}]{Huebner2020}%
  \BibitemOpen
  \bibfield  {author} {\bibinfo {author} {\bibfnamefont {M.}~\bibnamefont
  {{H{\"u}bner}}}, \bibinfo {author} {\bibfnamefont {C.}~\bibnamefont
  {{Talbot}}}, \bibinfo {author} {\bibfnamefont {P.~D.}\ \bibnamefont
  {{Lasky}}}, \ and\ \bibinfo {author} {\bibfnamefont {E.}~\bibnamefont
  {{Thrane}}},\ }\href {\doibase 10.1103/PhysRevD.101.023011} {\bibfield
  {journal} {\bibinfo  {journal} {\prd}\ }\textbf {\bibinfo {volume} {101}},\
  \bibinfo {eid} {023011} (\bibinfo {year} {2020})},\ \Eprint
  {http://arxiv.org/abs/1911.12496} {arXiv:1911.12496 [astro-ph.HE]}
  \BibitemShut {NoStop}%
\bibitem [{\citenamefont {Sellentin}\ and\ \citenamefont
  {Heavens}(2016)}]{Sellentin2016}%
  \BibitemOpen
  \bibfield  {author} {\bibinfo {author} {\bibfnamefont {E.}~\bibnamefont
  {Sellentin}}\ and\ \bibinfo {author} {\bibfnamefont {A.~F.}\ \bibnamefont
  {Heavens}},\ }\href {\doibase 10.1093/mnrasl/slv190} {\bibfield  {journal}
  {\bibinfo  {journal} {Mon. Not. R. Astron. Soc. Lett.}\ }\textbf {\bibinfo
  {volume} {456}},\ \bibinfo {pages} {L132} (\bibinfo {year} {2016})},\ \Eprint
  {http://arxiv.org/abs/1511.05969} {arXiv:1511.05969} \BibitemShut {NoStop}%
\bibitem [{\citenamefont {{Abbott}}\ \emph {et~al.}(2018)\citenamefont
  {{Abbott}} \emph {et~al.}}]{ObservingScenarios}%
  \BibitemOpen
  \bibfield  {author} {\bibinfo {author} {\bibfnamefont {B.~P.}\ \bibnamefont
  {{Abbott}}} \emph {et~al.},\ }\href {\doibase 10.1007/s41114-018-0012-9}
  {\bibfield  {journal} {\bibinfo  {journal} {Living Reviews in Relativity}\
  }\textbf {\bibinfo {volume} {21}},\ \bibinfo {eid} {3} (\bibinfo {year}
  {2018})},\ \Eprint {http://arxiv.org/abs/1304.0670} {arXiv:1304.0670 [gr-qc]}
  \BibitemShut {NoStop}%
\bibitem [{\citenamefont {{Aasi}}\ \emph {et~al.}(2015)\citenamefont {{Aasi}}
  \emph {et~al.}}]{AdvancedLIGO}%
  \BibitemOpen
  \bibfield  {author} {\bibinfo {author} {\bibfnamefont {J.}~\bibnamefont
  {{Aasi}}} \emph {et~al.},\ }\href {\doibase 10.1088/0264-9381/32/7/074001}
  {\bibfield  {journal} {\bibinfo  {journal} {Classical and Quantum Gravity}\
  }\textbf {\bibinfo {volume} {32}},\ \bibinfo {eid} {074001} (\bibinfo {year}
  {2015})},\ \Eprint {http://arxiv.org/abs/1411.4547} {arXiv:1411.4547 [gr-qc]}
  \BibitemShut {NoStop}%
\bibitem [{\citenamefont {{Abbott}}\ \emph {et~al.}(2016)\citenamefont
  {{Abbott}} \emph {et~al.}}]{O1BBH}%
  \BibitemOpen
  \bibfield  {author} {\bibinfo {author} {\bibfnamefont {B.~P.}\ \bibnamefont
  {{Abbott}}} \emph {et~al.},\ }\href {\doibase 10.1103/PhysRevX.6.041015}
  {\bibfield  {journal} {\bibinfo  {journal} {Physical Review X}\ }\textbf
  {\bibinfo {volume} {6}},\ \bibinfo {eid} {041015} (\bibinfo {year} {2016})},\
  \Eprint {http://arxiv.org/abs/1606.04856} {arXiv:1606.04856 [gr-qc]}
  \BibitemShut {NoStop}%
\bibitem [{\citenamefont {Vallisneri}\ \emph {et~al.}(2015)\citenamefont
  {Vallisneri}, \citenamefont {Kanner}, \citenamefont {Williams}, \citenamefont
  {Weinstein},\ and\ \citenamefont {Stephens}}]{Vallisneri2015}%
  \BibitemOpen
  \bibfield  {author} {\bibinfo {author} {\bibfnamefont {M.}~\bibnamefont
  {Vallisneri}}, \bibinfo {author} {\bibfnamefont {J.}~\bibnamefont {Kanner}},
  \bibinfo {author} {\bibfnamefont {R.}~\bibnamefont {Williams}}, \bibinfo
  {author} {\bibfnamefont {A.}~\bibnamefont {Weinstein}}, \ and\ \bibinfo
  {author} {\bibfnamefont {B.}~\bibnamefont {Stephens}},\ }\href {\doibase
  10.1088/1742-6596/610/1/012021} {\bibfield  {journal} {\bibinfo  {journal}
  {J. Phys. Conf. Ser.}\ }\textbf {\bibinfo {volume} {610}},\ \bibinfo {pages}
  {012021} (\bibinfo {year} {2015})},\ \Eprint {http://arxiv.org/abs/1410.4839}
  {arXiv:1410.4839} \BibitemShut {NoStop}%
\bibitem [{\citenamefont {{The LIGO Scientific Collaboration}}\ \emph
  {et~al.}(2019)\citenamefont {{The LIGO Scientific Collaboration}},
  \citenamefont {{the Virgo Collaboration and}~{Abbott}} \emph
  {et~al.}}]{OpenData}%
  \BibitemOpen
  \bibfield  {author} {\bibinfo {author} {\bibnamefont {{The LIGO Scientific
  Collaboration}}}, \bibinfo {author} {\bibfnamefont {R.}~\bibnamefont {{the
  Virgo Collaboration and}~{Abbott}}},  \emph {et~al.},\ }\href@noop {} {\
  (\bibinfo {year} {2019})},\ \Eprint {http://arxiv.org/abs/1912.11716}
  {arXiv:1912.11716 [gr-qc]} \BibitemShut {NoStop}%
\end{thebibliography}%

\appendix

\end{document}